\documentclass[superscriptaddress,onecolumn]{revtex4}
\usepackage{graphicx,epsfig}
\usepackage{amsmath}
\usepackage {amssymb}
\usepackage{multirow}
\usepackage{float}

\usepackage {ulem}

\usepackage[dvipsnames]{xcolor}
\RequirePackage[colorlinks,citecolor=blue,urlcolor=blue,linkcolor=blue]{hyperref}

\newcommand{\be}{\begin{eqnarray}}
\newcommand{\ee}{\end{eqnarray}}
\newcommand{\bea}{\begin{eqnarray}}

\newcommand{\eea}{\end{eqnarray}}

\makeindex

\begin{document}

%\begin{titlepage}

%\rightline{hep-th/yymmnnn}

%\vskip 2cm
\begin{center}
\large \bf {Quasinormal modes and bound states of massive scalar fields in wormhole spacetimes}
\end{center}
%\centerline{\large \bf {  using the WBK method }}

%\centerline{\large \bf {five-dimensions}}

\author{Sebasti\'{a}n Alfaro}
\email{sebastian.alfaroz@userena.cl}
\affiliation{Departamento de F\'isica, Facultad de Ciencias, Universidad de La Serena,\\
Avenida Cisternas 1200, La Serena, Chile.}

\author{P. A. Gonz\'{a}lez}
\email{pablo.gonzalez@udp.cl} \affiliation{Facultad de
Ingenier\'{i}a y Ciencias, Universidad Diego Portales, Avenida Ej\'{e}rcito
Libertador 441, Casilla 298-V, Santiago, Chile.}

\author{Diego Olmos}
\email{diego.olmos@userena.cl}
\affiliation{Departamento de F\'isica, Facultad de Ciencias, Universidad de La Serena,\\
Avenida Cisternas 1200, La Serena, Chile.}

\author{Eleftherios Papantonopoulos}
\email{lpapa@central.ntua.gr}
\affiliation{Physics Division, School of Applied Mathematical and Physical Sciences,
 National Technical University of Athens, 15780 Zografou Campus,
    Athens, Greece.}

\author{Yerko V\'asquez}
\email{yvasquez@userena.cl}
\affiliation{Departamento de F\'isica, Facultad de Ciencias, Universidad de La Serena,\\
Avenida Cisternas 1200, La Serena, Chile.}

%\vskip 1.0cm

\begin{abstract}

In this work we explore the propagation of massive scalar fields on some wormhole backgrounds. On one side, we consider the Bronnikov-Ellis wormhole solution and wormhole geometries with a non-constant redshift function by introducing a gravitational mass $M$, which goes over into the Bronnikov-Ellis wormhole when the gravitational mass parameter vanishes. We employ the continued fraction method to calculate accurately the quasinormal frequencies of massive scalar fields, particularly focusing on low values of the angular number, and we show an anomalous behaviour of the decay rate of the quasinormal frequencies, for $n \geq \ell$. Also, we show that for a massive scalar field and $M \neq 0$ the effective potential allows potential wells for some values of the parameters which support bound states, which are obtained using the continued fraction method and they are characterized by having only a frequency of oscillation and they do not decay; however, for the Bronnikov-Ellis wormhole the effective potential do not support bound states.
On the other side, we consider a wormhole geometry which is an exact solution of $f(R)$ modified gravity. For this geometry the quasinormal frequencies can be obtained analytically, being the longest-lived modes the ones with lowest angular number $\ell$. So, in this wormhole background the anomalous behaviour is avoided.

\end{abstract}

\maketitle

\tableofcontents

%\end{titlepage}

%\clearpage

%%%%%%%%%%%%%%%%%%%%%%%%
\section{Introduction}
%%%%%%%%%%%%%%%%%%%%%%%%

Solutions of Einstein equations that connect different parts of the Universe or two different Universes can be found in General Relativity (GR) known as wormholes. The concept of the wormhole can  be traced back to Flamm in 1916 \cite{Flamm} and then wormhole-type solutions were found  by Einstein and Rosen (ER), where they considered a physical space representing a physical space which was connected by a wormhole-type solution known as ER bridge \cite{Rosen}. Then, the pioneering work of Misner and Wheeler \cite{MisWheel} and Wheeler \cite{Wheel} developed further the  wormholes concept.
Lorentzian wormholes were studied in GR \cite{Bronnikov:1973fh,Ellis:1973yv,Morris,phantom,Visser}, in which conditions for traversable wormholes were found introducing a  static spherically symmetric metric. However, a condition on the wormhole throat leads to the violation of the null energy condition (NEC). To allow the formation of traversable wormhole geometries a matter distribution of exotic or phantom matter has to be introduced \cite{phantom}.

Recently there are many efforts to understand modifications of GR which are generated by the presence of a scalar field coupled to gravity. This coupling has important implications in local and global solutions in these theories known as scalar-tensor theories \cite{Fujii:2003pa}. One of the best studied    scalar-tensor theories is provided by the Horndeski Lagrangian \cite{Horndeski:1974wa}. The Horndeski theory has been studied in short and large distances. In this theory  black hole solutions were found \cite{Kolyvaris:2011fk,Rinaldi:2012vy,Kolyvaris:2013zfa,Babichev:2013cya,Charmousis:2014zaa}, known as  Galilean black holes, and also wormhole
geometries were generated \cite{Korolev:2014hwa}-\cite{Bakopoulos:2023tso}.

A study of Galileon black holes was performed in  \cite{Koutsoumbas:2018gbd}. The Regge-Wheeler potential generated by a test wave in the vicinity of a Galileon black hole potential was studied and it was investigated the formation and the behaviour of bound states
trapped in this potential well or penetrating the horizon of the Galileon black hole. The formation of the bound states was depending of how strongly matter is coupled to gravity expressed by the coupling the scalar field to Einstein tensor. In \cite{Chatzifotis:2020oqr} bound states in a wormhole geometry in the scalar-tensor Horndeski theory were studied in which the wormhole throat connects two Anti-de Sitter spacetimes. The scalar field is coupled kinetically to curvature and exact static spherically symmetric wormhole solutions were found. Using the Wentzel-Kramers-Brillouin (WKB) approximation the formation and propagation of bound states in the vicinity of a wormhole was studied. The behaviour of the bound states trapped in the potential wells and  the flow between the two AdS regions was investigated.

An important issue of the compact objects formed by  black holes or  wormholes is their stability. The relativistic collision of two of these compact objects produces gravitational waves which will help us to understand better  the gravitational interaction. However, the recent LIGO detections~\cite{Abbott:2016blz}-\cite{TheLIGOScientific:2017qsa} do not yet probe the detailed structure of spacetime beyond the photon sphere. Future observations may detect the ringdown phase, which is governed by a series of damped oscillatory modes at early times, which are known as quasinormal modes (QNMs) \cite{Vishveshwara:1970zz}-\cite{Cardoso:2016rao}.
Future gravitational observations may provide  information on different compact objects than black holes, objects without event horizons which are known as exotic compact objects (ECOs) \cite{Mazur:2001fv}-\cite{Holdom:2016nek}. Perturbations of black holes and wormholes were extensively studied in the literature. In the case that a test scalar perturbs black hole compact objects the partially reflected waves from the photon sphere (PS)  mirror off the AdS boundary and re-perturb the PS to give rise to a damped beating pattern. In the case of a wormhole  do not decay with time but have constant and equal amplitude to that of the initial ringdown. Therefore, the knowledge of QNMs and quasinormal frequencies (QNFs) can give us important information on the properties of compact objects and distinguish their nature. The QNMs give an infinite discrete spectrum which consists of complex frequencies, $\omega=\omega_R+i \omega_I$, where the real part $\omega_R$ determines the oscillation timescale of the modes, while the complex part $\omega_I$ determines their exponential decaying timescale (for a review on QNMs see \cite{Kokkotas:1999bd, Berti:2009kk}).

In the case that the probe scalar field is massive, a new mass scale is introduced and then a different behaviour  was found
\cite{Konoplya:2004wg, Konoplya:2006br,Dolan:2007mj, Tattersall:2018nve},  at least for the overtone $n = 0$. If the mass of the scalar field is  light, then the longest-lived QNMs are those with a high angular number $\ell$, whereas if the mass of the scalar field is large  the longest-lived modes are those with a low angular number $\ell$. One is expecting this behaviour because the fluctuations of the probe massive scalar field can maintain the QNMs to live longer even if the angular number $\ell$ is large, introducing in this way an  anomaly of the decaying modes depending on the mass of the scalar field exceeding  a critical value or not. Also, if a cosmological constant is introduced an anomalous behaviour of QNMs was found in \cite{Aragon:2020tvq} because there is interplay of the mass of the scalar field and the value of the cosmological constant. If the background metric is the Reissner-Nordstr\"om and the probe scalar field is massless \cite{Fontana:2020syy} of massive \cite{Gonzalez:2022upu} depending on critical values of the charge of the black hole, the charge of the scalar field and its mass an anomalous behaviour was also found. The decay modes of QNMs in various setups were studied in \cite{Aragon:2020teq}-\cite{Becar:2023zbl}.

QNMs and QNFs have also been calculated in the background of wormholes. The perturbations of wormhole spacetime was investigated in \cite{Konoplya:2005et}. It was found that the QNMs behave the same way as in black hole background and  includes the quasinormal ringing and power-law asymptotically late time tails. However, in \cite{Konoplya:2016hmd} it was found that the symmetric WHs do not have the same ringing behaviour of the BHs  at a few various dominant multipoles. In the case of massive probe scalar field perturbations,    it was shown \cite{Churilova:2019qph} that wormholes  with a constant redshift function do not allow for  longest-lived modes, while wormholes with non-vanishing radial tidal force do allow for quasi-resonances. Calculating the high frequency QNMs in the case of the Morris-Thorne wormhole spacetime, the shape function of a spherically symmetric traversable Lorenzian wormhole near its throat was constructed in \cite{Konoplya:2018ala}. Scalar and axial QNMs of massive static phantom WHs were calculated in \cite{Blazquez-Salcedo:2018ipc}.

In \cite{Gonzalez:2022ote} it was shown that if the probe scalar field is massive then an anomalous behaviour of QNMs was also observed if the background metric defines a wormhole geometry. It was shown that in the case of  Bronnikov-Ellis wormhole \cite{Ellis:1973yv}  there is a critical mass of the scalar field beyond which the anomalous decay rate of the QNMs is present. In the case of the  Morris-Thorne wormhole \cite{Morris:1988tu} the anomalous decay rate of the QNMs is not present at least for the fundamental modes.  In \cite{Azad:2022qqn} polar perturbations of static Bronnikov-Ellis wormholes were studied  and the QNMs of rapidly rotating Bronnikov-Ellis wormholes were recently analyzed in \cite{Khoo:2024yeh}.

As we already discussed in the case of the Horndeski scalar-tensor theory if matter, parameterized by a phantom scalar field, is strongly coupled to gravity then  bound states  are formed \cite{Koutsoumbas:2018gbd,Chatzifotis:2020oqr}. In this work we follow a much simpler approach. We consider the Bronnikov-Ellis wormhole solution \cite{Bronnikov:1973fh,Ellis:1973yv} and wormhole geometries with non-constant redshift function by introducing a gravitational mass $M$, which goes over into the Bronnikov-Ellis wormhole when $M=0$.
We perturb the massive wormhole background by a test massive scalar field.  Using the continued fraction method we calculate accurately the QNFs for low values of the angular number. We find stability of the background wormhole and calculating the effective potential we show that for an interplay of the mass parameter and the phantom scalar charge,  potential wells are generated  supporting bound states.

We also study the QNFs in a wormhole spacetime supported by a phantom scalar field in which the scalar curvature is generalized to an $f(R)$ function \cite{Karakasis:2021lnq,Karakasis:2021rpn,Tang:2019qiy,DeFalco:2021ksd,Karakasis:2021tqx}. In this model the scalar field self-interacts
with a mass term potential which is derived from the scalar equation and in the resulting $f(R)$ model
the scalar curvature is modified by the presence of the scalar field and it is free of ghosts and avoids
the tachyonic instability.

The work is organized as follows. In Section~\ref{sec2} we review the solutions of the  traversable wormholes. In Section~\ref{sec3}  we study the perturbations of massive scalar field in the wormhole geometry. In Section~\ref{sec4} we calculate the QNFs of the non-constant redshift function wormholes, resulting in a non-zero tidal radial force, and the Bronnikov-Ellis wormhole numerically using the continued fraction method. Also, we show that bound states are allowed by the wormholes with non-vanishing radial tidal force and the frequencies of the lowest bound states are calculated using the continued fraction method. In Section~\ref{sec5} we calculate the QNFs of an $f(R)$ gravity wormhole analytically and using the WKB method. In Section~\ref{sec6} are our conclusions.

\section{Review of  traversable wormholes}

\label{sec2}

Morris and Thorne \cite{Morris} were the first they found the necessary conditions, for a  static spherically symmetric metric, to generate traversable Lorentzian wormholes as  exact solutions in GR. To find a solution, conditions on the wormhole throat necessitate the introduction of   exotic matter, which   leads to the violation of the NEC.

We consider the following action
\begin{equation} S = \int d^4x \sqrt{-g} \left( \frac{R}{2} + \frac{1}{2}\partial^{\mu}\phi\partial_{\mu}\phi -V(\phi) \right) ~,\label{action} \end{equation}
which consists of  the Ricci scalar $R$, a scalar field with negative kinetic energy and a self-interacting potential. The field equations that emerge through the variation of this action  are
\begin{eqnarray}
R_{\mu\nu} - \frac{1}{2}g_{\mu\nu}R  &=& T_{\mu\nu}^{\phi}~, \label{EE}\\
\Box \phi + V_{\phi}(\phi) &=&0~, \label{KG}
\end{eqnarray}
where $\Box = \nabla^{\mu}\nabla_{\mu}$ represents the D'Alambert operator with respect to the metric, and the energy-momentum tensor is given by
\begin{equation} T_{\mu\nu}^{\phi} = - \partial_{\mu}\phi \partial_{\nu}\phi + \frac{1}{2}g_{\mu\nu}\partial^{\alpha}\phi \partial_{\alpha}\phi - g_{\mu\nu}V(\phi)~. \end{equation}
We consider the metric ansatz proposed by  Morris and Thorne \cite{Morris} in spherical coordinates, which is given by
\begin{equation} ds^2 = - e^{2\Phi(r)}dt^2 + \left(1-\frac{b(r)}{r}\right)^{-1} dr^2 + r^2 d\Omega\label{MorrisThorne-metric} ~,\end{equation}
where $\Phi(r)$ is the redshift function, $b(r)$ is the shape function and $d\Omega = d\theta^2 + \sin(\theta)^2d\varphi^2$. To achieve  a wormhole geometry, it is necessary for these functions to adhere to the following conditions \cite{Morris,Visser}
\begin{enumerate}
    \item $\frac{b(r)}{r}\leq 1$ must hold for every $[r_{th},+\infty)$, where $r_{th}$ denotes the radius of the throat. This condition ensures the finiteness of the proper radial distance, defined by $l(r)=\pm\int_{r_{th}}^{r}\left(1-\frac{b(r)}{r}\right)^{-1} dr$, across the entire spacetime.
   Note that in the coordinates $(t,l,\theta,\varphi)$ the line element \eqref{MorrisThorne-metric} can be expressed as \begin{align}
    ds^2=-e^{2\Phi(l)}dt^2+dl^2+r^2(l)(d\theta^2+\sin^2\theta d\varphi)\,.\label{metric-proper-chart}
    \end{align}
   Here, the throat radius corresponds to $r_{th}=\min\{r(l)\}$.
    \item $\frac{b(r_{th})}{r_{th}}=1$ at the throat. This condition arises from the requirement that the throat constitutes a stationary point of $r(l)$. Alternatively, it can be derived from the demand that the embedded surface of the wormhole be vertical at the throat.
    \item $b'(r)<\frac{b(r)}{r}$ which reduces to $b'(r_{th})\leq 1$ at the throat. This condition, known as the flare-out condition, guarantees that $r_{th}$ is indeed a minimum and not any other type of stationary point.
    \end{enumerate}

To ensure the absence of horizons and singularities requires  $\Phi(r)\neq 0$  which means that $\Phi(r)$ is finite throughout the spacetime. The Ellis \cite{Ellis:1973yv} is a wormhole solution of an action that consists of a pure Einstein-Hilbert term and a scalar field with negative kinetic energy
\begin{equation}
	S = \int d^4x \sqrt{-g} \left(\frac{R}{2} + \frac{1}{2}\partial^{\mu}\phi\partial_{\mu}\phi\right)~.
\end{equation}
Given the metric ansatz (\ref{MorrisThorne-metric}) and setting $V(r)=0$,  the aforementioned equations yield the following solution
\begin{eqnarray}
\Phi (r) &=& 0 \,, \\
b(r) &=&A^2/r~, \label{bellis}\\
\phi(r) &=& \sqrt{2} \tan ^{-1}\left(\frac{\sqrt{r^2-A^2}}{A}\right)+\phi_{\infty}~, \label{ellisscalar}\\
R(r) &=& -\frac{2 A^2}{r^4}~, \label{rellis}
\end{eqnarray}
where $A$, $\phi_{\infty}$ are constants of integration. The  spacetime is asymptotically flat, which is evident in the behaviour of $b(r)$ and $R(r)$. The wormhole throat is determined by the solution of the equation
\begin{equation} g_{rr}^{-1} = 0 \rightarrow r_{th} = \pm A~. \end{equation}
Additionally, the solution satisfies the flare-out condition, and it meets the requirement $b'(r_{th})=-1$.
%\newline

A constant value at infinity $\phi(r\to \infty) = \frac{\pi }{\sqrt{2}} +\phi_{\infty} $ has  scalar field takes, so to make the scalar field vanish at large distances we will set $\phi_{\infty} = -\frac{\pi }{\sqrt{2}}$.  The scalar field takes the asymptotic value at infinity at the position of the throat $\phi(r=r_{th})=\phi_{\infty}$.  From  the solution one  can be see in (\ref{bellis}), (\ref{ellisscalar}) and (\ref{rellis}) the integration constant $A$  of the phantom scalar field plays a very important role in the formation of the wormhole geometry.  It has  units $[L]^2$, appears in the scalar curvature and at the position of the throat takes the value $R_{r = r_{\text{th}}}(A) = -2/A^2$. Also it effects the size of the throat, a larger charge $A$ gives a larger wormhole throat. The above discussion indicates that  the phantom scalar field is very important for the generation of the wormhole geometry and to define the scalar curvature and specifying  the throat of the wormhole geometry.

The function $b(r)$ encodes the shape of the wormhole and at certain minimum value of $r$, the throat of a wormhole is defined, namely, when  $r_{min} = b_0$. Thus,  the radial coordinate increases ranging from $r_{min}$ until spatial infinity $r = \infty$. Also  $\Phi(r)$ must be finite everywhere in light of the requirement of absence of singularities. In addition, $\Phi(r) \rightarrow 0$ as $r \rightarrow \infty $ (or $ l \rightarrow \pm \infty$) based on the requirement of asymptotic flatness.
On the other hand, the shape function $b(r)$ should satisfy that $1- b(r)/r > 0$ and
$b(r)/r \rightarrow 0$ as $r \rightarrow \infty $ (or equivalently $ l \rightarrow \pm \infty$).
In the throat $r = b(r)$ and thus $1- b(r)/r$ vanishes.
Traversable wormholes does not have a singularity in the throat. The later means that travellers can pass through the wormhole during the finite time.

We will consider the following metric functions
\begin{equation} \label{wormhole}
    e^{2 \Phi(r)} = 1 -\frac{2M}{r}\,, \,\,\,\, b(r)= \frac{b_0^2}{r} \,,
\end{equation}
which corresponds to the Bronnikov-Ellis wormhole in the limit $M \rightarrow 0$ \cite{Bronnikov:1973fh,Ellis:1973yv}. Note that the parameter $b_0$ specifies the charge of the phantom scalar field.
%\newline

The most general energy-momentum tensor compatible with wormhole geometries is given by \cite{Bronnikov:2018vbs}
\begin{equation}
    T^{\mu}_{\,\,\,\,\, \nu} = diag(- \rho, p_r, p_t, p_t) \,,
\end{equation}
where $\rho$ denotes the energy density, and $p_r$ and $p_t$ represent the radial and tangential pressures, respectively.

The Einstein field equations take the form
\begin{eqnarray}
\nonumber \rho &=& \frac{b'(r)}{r^2} \,,  \\
\nonumber p_r &=& \frac{-b(r)+2 r (r-b(r)) \Phi '(r)}{r^3} \,,  \\
\nonumber p_t &=& \frac{(b(r)- r b'(r) +2 r (r-b(r)) \Phi'(r))(1+ r \Phi'(r)) + 2 (r - b(r))r^2 \Phi''(r)}{2 r^3} \,.
\end{eqnarray}
The important energy conditions include the Weak Energy Condition (WEC), the Null Energy Condition (NEC), the Strong Energy Condition (SEC) and the Dominant Energy Condition (DEC). These conditions, expressed in terms of the principal pressures, are given by, see for instance \cite{Samanta:2019tjb}
\begin{eqnarray}
  \nonumber  && \text{WEC}: \rho \geq 0, \rho + p_r \geq 0, \rho + p_t \geq 0 \\
  \nonumber  && \text{NEC}: \rho + p_r \geq 0, \rho+ p_t \geq 0 \\
   \nonumber && \text{SEC}: \rho+ p_r \geq 0, \rho+ p_t \geq 0, \rho+ p_r+ 2 p_t \geq 0 \\
  \nonumber  && \text{DEC}: \rho \geq 0, \rho - |p_r| \geq 0, \rho - |p_t| \geq 0
\end{eqnarray}
For the wormhole geometry (\ref{wormhole}), the obtained expressions are as follows
\begin{eqnarray}
 \nonumber   \rho &=& -\frac{b_0^2}{r^4}\,,  \,\,\,\,\,  p_r = \frac{-b_0^2 + 2M r}{r^3(r-2M)} \,, \,\,\,\,\,  p_t = \frac{(b_0^2 (M-r)+Mr^2)}{r^4 (r- 2M)^2} \,, \\
  \nonumber  \rho+ p_r &=& \frac{2(b_0^2(M-r)+ M r^2)}{r^4 (r- 2M)} \,, \,\,\,\,\,  \rho+ p_t = \frac{M (b_0^2 (2 r -3M)+ (M-r)r^2)}{r^4 (r-2M)^2}\,.
\end{eqnarray}
Hence, $\rho < 0$, leading to the violation of the Weak Energy Condition for any $r > r_{th} = b_0$. Furthermore, in the range $b_0 \leq r < \frac{b_0^2+\sqrt{b_0^4- 4b_0^2 M^2}}{2 M}$, $\rho + p_r$ is negative. Additionally, $p_t$ becomes negative for $r > \frac{b_0^2+\sqrt{b_0^4- 4b_0^2 M^2}}{2 M}$. Consequently, this implies that $\rho+ p_t$ is at least negative in that range. Consequently, all the energy conditions are violated for $r > r_{th}$.

\section{Massive scalar field perturbations}
\label{sec3}

To investigate the propagation of a massive scalar field in a wormhole geometry, we consider  the Klein-Gordon equation, which is expressed as
\begin{equation}
\frac{1}{\sqrt{-g}} \partial_\mu (\sqrt{-g} g^{\mu \nu} \partial_\nu) \Psi = m^2 \Psi\,.
\end{equation}
To decouple and subsequently solve the Klein-Gordon equation, we employ the method of separation of variables, utilizing the following ansatz
\begin{equation}\label{separable}
\Psi(t,r,\theta,\varphi) = e^{-i \omega t} \: \frac{u(r)}{r} \: Y_{\ell} (\Omega)\,.
\end{equation}
Here, $\omega$ represents the unknown frequency (to be determined), and $Y_{\ell}(\Omega)$ denotes the spherical harmonics, dependent solely on the angular coordinates \cite{book}.

After implementing the aforementioned mentioned ansatz, it becomes straightforward to derive  a Schr{\"o}dinger-like equation  for the radial part, specifically
\begin{equation}
\frac{\mathrm{d}^2 u}{\mathrm{d}r^{\ast}{}^2} + [ \omega^2 - V(r^{\ast}) ] u = 0\,,
\label{sequation1}
\end{equation}
where $r^{\ast}$ is the well-known tortoise coordinate, i.e.,
\begin{equation}
r^{\ast}  \equiv  \int \frac{\mathrm{d}r}{e^{\Phi} \sqrt{1 -\frac{b(r)}{r}}}\,.
\end{equation}
Finally, the effective potential for scalar field perturbations is given by
\begin{equation}
V(r) = e^{2\Phi} \: \left(m^2 + \frac{\ell (\ell+1)}{r^2}
-
\frac{b' r - b}{2r^3}
+
\frac{1}{r}
 \Bigg(1 -\frac{b}{r}\Bigg) \Phi'
 \right)\,,
\label{pot}
\end{equation}
where, the prime denotes differentiation with respect to the radial coordinate, and $\ell  \geq 0$ is the angular degree. Therefore, we have transformed the problem into the well-known one-dimensional Schr{\"o}dinger equation with energy $\omega^2$ and an effective potential $V(r)$.

\section{Bronnikov-Ellis wormholes and non-constant redshift function wormholes}
\label{sec4}

\begin{figure}[h!]
\begin{center}
\includegraphics[width=0.5\textwidth]{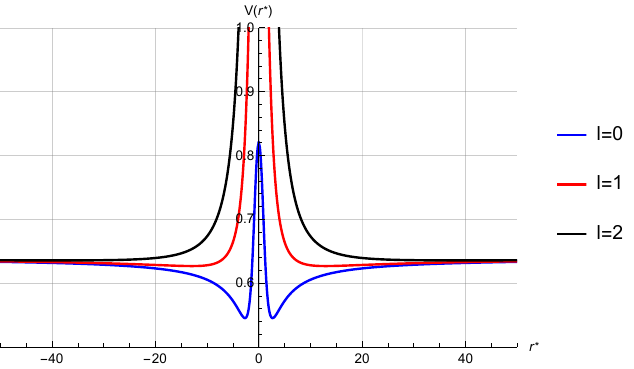}\\
\end{center}
\caption{Effective potential of the  wormhole spacetime (\ref{wormhole}), for $M/b_0 = 0.25$, $m b_0 = 0.8$ and $\ell = 0, 1, 2$.}
\label{worm}
\end{figure}

We consider the following metric functions
\begin{equation}
    e^{2 \Phi(r)} = 1 -\frac{2M}{r}\,, \,\,\,\, b(r)= \frac{b_0^2}{r} \,,
\end{equation}
which goes over into the Bronnikov-Ellis wormhole in the limit $M \rightarrow 0$. Thus, the effective potential of scalar perturbations is given by
\begin{equation}
V(r) = \frac{-3 M q^2 + q^2 r + (1-2 \ell (\ell+1))M r^2 + \ell (\ell+1)r^3}{r^5} + \left( 1- \frac{2M}{r} \right) m^2 \,,
\end{equation}
which is plotted in Fig. \ref{worm}, where a barrier-like shape is observed. Additionally, a well may form near the throat for a massive scalar field, $M \neq 0$ and at low values of $\ell$, while for high values of $\ell$, the well disappears; therefore, the potential can support bound states for low values of $\ell$.

For $M=0$ the effective potential transforms into $V(r)= q^2/r^4+ \ell (\ell+1)/r^2 + m^2$, and it is straightforward to demonstrate that it cannot give rise to a well; hence, bound states are prohibited in this case.
\newline

On the other hand, Eq. (\ref{sequation1}) can be written as
\begin{eqnarray}
\label{nueva}
\nonumber   &&  (2 M-r) \left(b_0^2-r^2\right)u''(r)+  \frac{\left(-3 M b_0^2+M r^2+b_0^2 r\right)}{r}u'(r)+  \\
    && \frac{M \left(\left( 2 m^2 r^2 + 2 \ell^2+2 \ell-1\right) r^2+3 b_0^2\right)-r \left(r^2 \left(\ell^2+\ell+ r^2( m^2-\omega ^2)\right)+b_0^2\right)}{r^2} u(r) =0\,,
\end{eqnarray}
which we will solve numerically to determine both the QNFs and the frequencies of bound states.

\subsection{Quasinormal frequencies}

In this section we use the continued fraction method (CFM) to calculate accurately the QNFs for massive scalar fields propagating on the Bronnikov-Ellis wormhole and non-constant redshift function wormholes described by Eq. (\ref{wormhole}). The CFM was devised by Leaver to compute the QNMs of Schwarzschild and Kerr black holes \cite{Leaver:1985ax, Leaver:1990zz}, and improved later by Nollert \cite{Nollert:1993zz}. It has been used to computed the QNMs in several black holes backgrounds, see for instance  \cite{Cardoso:2003qd, Konoplya:2013rxa, Richartz:2014jla, Chowdhury:2018pre,Becar:2022wcj}. However, to our knowledge this is the first time that this method is applied to calculate the QNFs in wormhole geometries.

To solve the radial equation it is necessary to specify the boundary conditions. Given the asymptotically flat nature of the wormhole geometries and the fact that the effective potential forms a barrier approaching
$V(r^{\ast}) \rightarrow m^2$ as $r^{\ast} \rightarrow \pm \infty$, we will consider the boundary conditions that demand purely outgoing waves at both infinities. In other words, no waves are coming from either asymptotically flat region \cite{Konoplya:2005et,Bronnikov:2021liv}
\begin{align}
\nonumber  &\Psi \sim e^{i \sqrt{\omega^2-m^2} r^{\ast}}\,, \quad r^{\ast} \rightarrow + \infty\,.
\\
&\Psi \sim e^{ - i \sqrt{\omega^2-m^2} r^{\ast}}\,, \quad r^{\ast} \rightarrow - \infty\,.
\end{align}
However, for the CFM it is more convenient to implement boundary conditions at the throat  and at infinity taking into account the symmetry of the potential, as it is explained next.
Therefore, the boundary condition that there are only outgoing waves at infinity is satisfied by the radial solution
%QNMs
%at infinity is given by
\begin{equation}
\nonumber   u \sim e^{i \Omega r^{\ast}}   \,\,\,  \text{as} \,\,\,   r^{\ast} \rightarrow \infty\,,
\end{equation}
where $\Omega = \sqrt{\omega^2-m^2}$,
%So, there are only outgoing waves at infinity,
which can be rewritten as
%transformed to
\begin{equation} \label{bc}
u \sim  e^{i \Omega r} r^{i M \omega^2/\Omega} \,\,\,    \text{as} \,\,\, r \rightarrow \infty\,.
\end{equation}
Also, since the potential is symmetric about the throat of the wormhole, which is located at $r_0^{\ast}=0$, the solutions will be symmetric or anti-symmetric. Therefore, we impose the following boundary conditions at the throat, $\frac{du}{d r^{\ast}}\left|_{r_0^{\ast}}=0\right.$ for the symmetric solutions and $u(r_0^{\ast}) = 0$ for the anti-symmetric solutions. These boundary conditions yields the even and odd overtones, respectively. It is worth mentioning that similar boundary conditions have been applied for other wormhole geometries, where the QNMs were obtained by direct integration of the wave equation \cite{Aneesh:2018hlp, DuttaRoy:2019hij}.
%\newline

It is convenient to consider  the following ansatz  for the symmetric solutions of the radial equation (\ref{nueva}) which satisfies the desired boundary conditions
\begin{equation}
u(r) = e^{i \Omega r} r^{i M \omega^2/\Omega} \sum_{n=0}^{\infty} a_n \left(
\frac{r-b_0}{r} \right)^n \,.
\end{equation}
So, substituting this expression in the radial equation, the following five-term recurrence relation is satisfied by the expansion coefficients
\begin{eqnarray} \label{rel}
\notag && c_0(0)a_1+c_1(0)a_0 =0\,, \\
\notag && c_0(1)a_2+c_1(1)a_1+c_2(1)a_0 =0\,, \\
\notag && c_0(2)a_3+c_1(2)a_2+c_2(2)a_1+c_3(2)a_0 =0 \,, \\
&& c_0(n)a_{n+1}+c_1(n)a_{n}+c_2(n)a_{n-1}+c_3(n)a_{n-2}+c_4(n)a_{n-3} = 0 \,, \,\, n \geq 3 \,,
\end{eqnarray}
where
\begin{eqnarray}
\notag c_0(n) &=& - (n+1) (2 n+1) (b_0-2 M) \left(m^2-\omega ^2\right) \,, \\
\notag c_1(n) &=& b_0^3 \left(m^2-\omega ^2\right)^2-i b_0^2 \left(m^2-\omega ^2\right) \left((4 n+1) \sqrt{\omega ^2-m^2}-2 i m^2 M\right)+b_0 \Bigg(m^2 \bigg(\ell^2+\ell+n \left(8 i M \sqrt{\omega ^2-m^2}+5 n+2\right) \\
\notag && + 2 i M \sqrt{\omega ^2-m^2}+1\bigg)-\omega ^2 \left(\ell^2+\ell+n \left(4 i M \sqrt{\omega ^2-m^2}+5 n+2\right)+i M \sqrt{\omega ^2-m^2}+1\right)\Bigg)+ \\
\notag  && 2 M \left(-\left(\ell^2+\ell+n (7 n+2)+1\right) \left(m^2-\omega ^2\right)-4 i M n \omega ^2 \sqrt{\omega ^2-m^2}-i M \omega ^2 \sqrt{\omega ^2-m^2}\right)  \,, \\
\notag c_2(n) &=& b_0 \Bigg(\omega ^2 \left(n \left(6 i M \sqrt{\omega ^2-m^2}+4 n-3\right)-4 i M \sqrt{\omega ^2-m^2}+6 M^2 \omega ^2+1\right)-m^2 \bigg(n \left(12 i M \sqrt{\omega ^2-m^2}+4 n-3\right)  \\
\notag &&  - 6 i M \sqrt{\omega ^2-m^2}+8 M^2 \omega ^2+1\Bigg)\bigg)+b_0^2 \left(m^2-\omega ^2\right) \left(2 m^2 M+i (2 n-1) \sqrt{\omega ^2-m^2}\right)+M \big((2 \ell (\ell+1)+n (18 n- \\
\notag && 17)+7) \left(m^2-\omega ^2\right)+20 i M n \omega ^2 \sqrt{\omega ^2-m^2}-8 i M \omega ^2 \sqrt{\omega ^2-m^2}+4 M^2 \omega ^4\big) \,,   \\
\notag  c_3(n) &=& b_0 \Bigg(m^2 \left(n \left(4 i M \sqrt{\omega ^2-m^2}+n-2\right)-5 i M \sqrt{\omega ^2-m^2}+4 M^2 \omega ^2+1\right)-\omega ^2 \bigg(n \left(2 i M \sqrt{\omega ^2-m^2}+n-2\right) \\
\notag && -3 i M \sqrt{\omega ^2-m^2}+3 M^2 \omega ^2+1\bigg)\Bigg)+M \bigg(-16 i M n \omega ^2 \sqrt{\omega ^2-m^2}+17 i M \omega ^2 \sqrt{\omega ^2-m^2}-(2 n (5 n-11)+13) \\
\notag && \left(m^2-\omega ^2\right)-6 M^2 \omega ^4\bigg) \,, \\
c_4(n) &=& M \left(m^2 (n-2) (2 n-3)+\omega ^2 \left(n \left(4 i M \sqrt{\omega ^2-m^2}-2 n+7\right)-7 i M \sqrt{\omega ^2-m^2}+2 M^2 \omega ^2-6\right)\right) \,.
\end{eqnarray}
%\newline
On the other hand, for the anti-symmetric solutions it is convenient to define
\begin{equation}
u(r) = (r-b_0)^{1/2} e^{i \Omega r} r^{i M \omega^2/\Omega -1/2} \sum_{n=0}^{\infty} a_n \left( \frac{r-b_0}{r} \right)^n \,,
\end{equation}
which incorporates the desired boundary conditions. Thus, substituting this expression in the radial equation, also a five-term recurrence relation (\ref{rel}) is satisfied by the expansion coefficients, but now the coefficients are given by
\begin{eqnarray}
\notag c_0(n) &=& -4 (n+1) (2 n+3) \left(b_0-2 M\right) \left(m^2-\omega ^2\right)  \,, \\
\notag c_1(n)&=& b_0 \Bigg(m^2 \left(4 \ell (\ell+1)+4 n \left(8 i M \sqrt{\omega ^2-m^2}+5 n+7\right)+24 i M \sqrt{\omega ^2-m^2}+13\right)-\omega ^2 \bigg(4 \ell (\ell+1)+ \\
\notag && 4 n \left(4 i M \sqrt{\omega ^2-m^2}+5 n+7\right)+12 i M \sqrt{\omega ^2-m^2}+13\bigg)\Bigg)-4 b_0^2 \left(m^2-\omega ^2\right) \left(2 m^2 M+i (4 n+3) \sqrt{\omega ^2-m^2}\right) + \\
\notag &&  4 b_0^3 \left(m^2-\omega ^2\right)^2+2 M \left(-(4 \ell (\ell+1)+4 n (7 n+9)+15) \left(m^2-\omega ^2\right)-16 i M n \omega ^2 \sqrt{\omega ^2-m^2}-12 i M \omega ^2 \sqrt{\omega ^2-m^2}\right) \,, \\
\notag c_2(n) &=& -2 b_0 m^2 \left(2 n \left(12 i M \sqrt{\omega ^2-m^2}+4 n+1\right)+16 M^2 \omega ^2+1\right)+2 b_0 \omega ^2 \bigg(2 n \left(6 i M \sqrt{\omega ^2-m^2}+4 n+1\right)-  \\
\notag && 2 i M \sqrt{\omega ^2-m^2}+12 M^2 \omega ^2+1\bigg)+8 b_0^2 \left(m^2-\omega ^2\right) \left(m^2 M+i n \sqrt{\omega ^2-m^2}\right)+4 M \bigg(\left(2 \ell (\ell+1)+18 n^2+n+3\right) \\
\notag &&  \left(m^2-\omega ^2\right)+20 i M n \omega ^2 \sqrt{\omega ^2-m^2}+2 i M \omega ^2 \sqrt{\omega ^2-m^2}+4 M^2 \omega ^4\bigg) \,, \\
\notag c_3(n) &=& b_0 \Bigg(m^2 \left(4 n \left(4 i M \sqrt{\omega ^2-m^2}+n-1\right)+4 M \left(4 M \omega ^2-3 i \sqrt{\omega ^2-m^2}\right)+1\right)-\omega ^2 \bigg(4 n \left(2 i M \sqrt{\omega ^2-m^2}+n-1\right)  \\
\notag && -8 i M \sqrt{\omega ^2-m^2}+12 M^2 \omega ^2+1\bigg)\Bigg)+2 M \bigg(-32 i M n \omega ^2 \sqrt{\omega ^2-m^2}+18 i M \omega ^2 \sqrt{\omega ^2-m^2}-(4 n (5 n-6)+9) \\
\notag && \left(m^2-\omega ^2\right)-12 M^2 \omega ^4\bigg) \,, \\
\notag c_4(n)&=& 4 M \left(m^2 (n-1) (2 n-3)+\omega ^2 \left(n \left(4 i M \sqrt{\omega ^2-m^2}-2 n+5\right)-5 i M \sqrt{\omega ^2-m^2}+2 M^2 \omega ^2-3\right)\right) \,.
\end{eqnarray}

For $M=0$, the coefficients $c_4(n)$ become zero for both symmetric and antisymmetric solutions; hence, the expansion coefficients satisfy four-term recurrence relations for the Bronnikov-Ellis wormhole.
Then, after applying Gaussian elimination twice for $M \neq 0$, and once for $M=0$, the original recurrence relations (\ref{rel}) can be simplified to a three-term relation, taking on the following concise form
\begin{eqnarray}
\notag \alpha_0 a_1 + \beta_0 a_0 &=& 0  \,, \\
\notag \alpha_n a_{n+1} + \beta_n a_n + \gamma_n a_{n-1} &=&  0 \,, \,\,\,\, n = 1, 2, ...
\end{eqnarray}
We skip providing the explicit expressions for $\alpha_n$, $\beta_n$ and $\gamma_n$ of the final three-term recurrence relations.
Now, according to Leaver \cite{Leaver:1985ax, Leaver:1990zz}, the recursion coefficients must satisfy the following continued fraction relation for the convergence of the series
\begin{equation}
\beta_0 - \frac{\alpha_0 \gamma_1}{\beta_1-}\frac{\alpha_1 \gamma_2}{\beta_2-} \cdots\frac{\alpha_n \gamma_{n+1}}{\beta_{n+1}-} \cdots = 0\,,
\end{equation}

\begin{figure}[h]
\begin{center}
\includegraphics[width=0.4\textwidth]{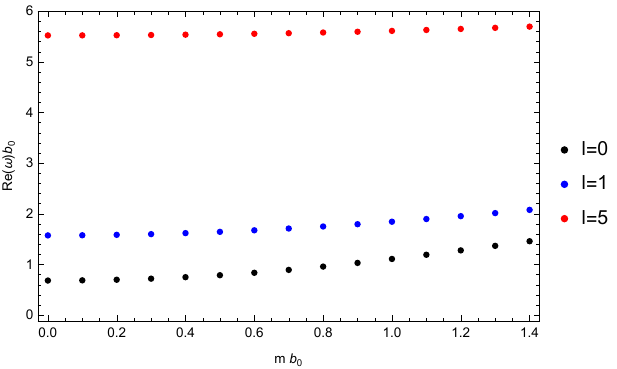}
\includegraphics[width=0.4\textwidth]{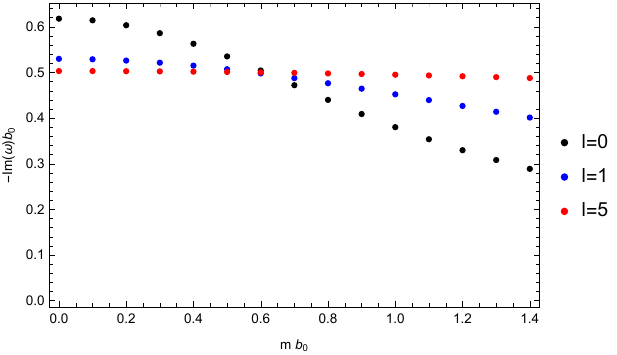}\\
\end{center}
\caption{The behaviour of $Re(\omega)b_0$ and $- Im(\omega)b_0$ for massive scalar field as a function of $m b_0$ for the Bronnikov-Ellis wormhole with $n=0$ (fundamental frequency), and $\ell = 0, 1, 5$ using the CFM.}
\label{omega_1}
\end{figure}

\begin{figure}[h]
\begin{center}
\includegraphics[width=0.4\textwidth]{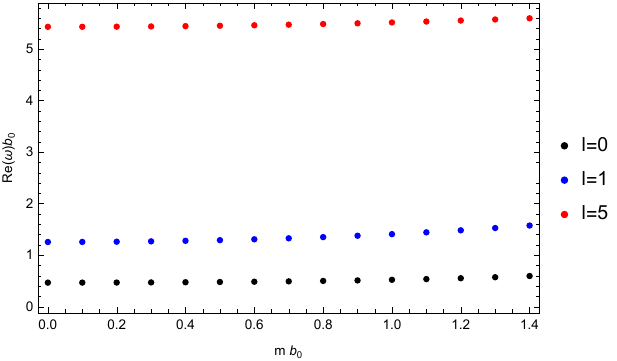}
\includegraphics[width=0.4\textwidth]{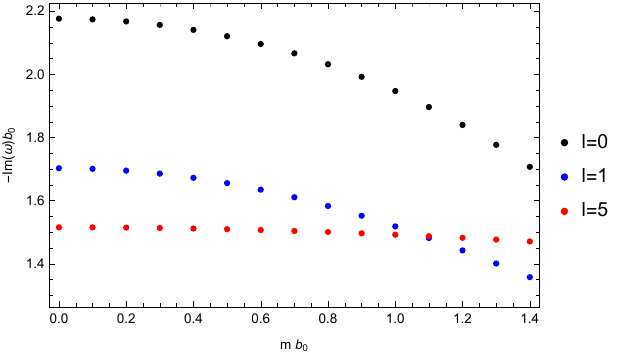}\\
\end{center}
\caption{The behaviour of $Re(\omega)b_0$ and $- Im(\omega)b_0$ for massive scalar field as a function of $m b_0$ for the Bronnikov-Ellis wormhole with $n=1$ (first overtone), and $\ell = 0, 1, 5$ using the CFM.}
\label{omega_4}
\end{figure}

\begin{figure}[h]
\begin{center}
\includegraphics[width=0.4\textwidth]{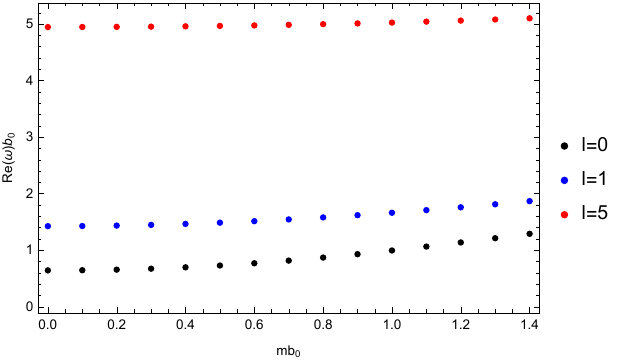}
\includegraphics[width=0.4\textwidth]{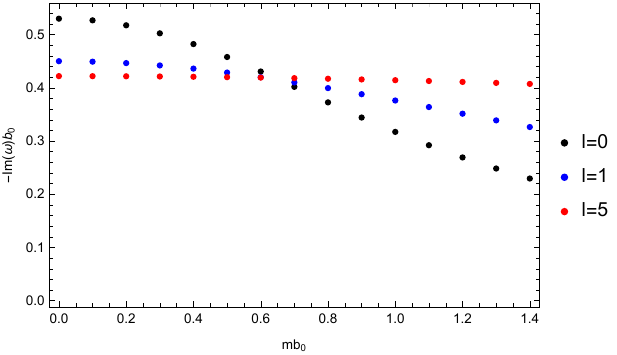}\\
\end{center}
\caption{The behaviour of $Re(\omega)b_0$ and $- Im(\omega)b_0$ for a massive scalar field as a function of $m b_0$ for the wormhole spacetime described by Eq. (\ref{wormhole}), with $M/b_0=0.1$, $n=0$ (fundamental frequency), and $\ell = 0, 1, 5$ using the CFM.}
\label{omega1}
\end{figure}

\begin{figure}[h]
\begin{center}
\includegraphics[width=0.4\textwidth]{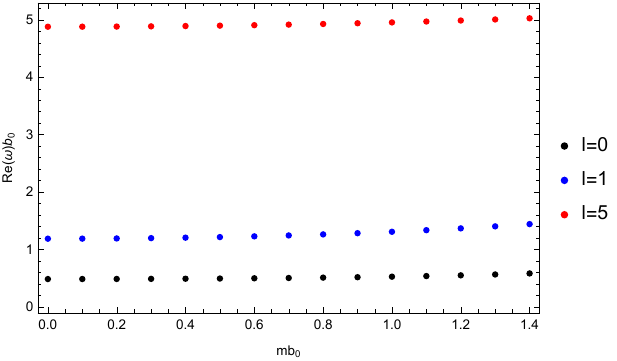}
\includegraphics[width=0.4\textwidth]{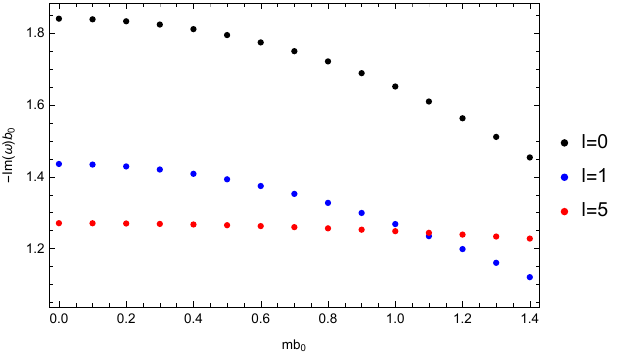}\\
\end{center}
\caption{The behaviour of $Re(\omega)b_0$ and $- Im(\omega)b_0$ for a massive scalar field as a function of $m b_0$ for the wormhole spacetime described by Eq. (\ref{wormhole}), with $M/b_0=0.1$, $n=1$ (first overtone), and $\ell = 0, 1, 5$ using the CFM.}
\label{omega2}
\end{figure}

\begin{figure}[h]
\begin{center}
\includegraphics[width=0.4\textwidth]{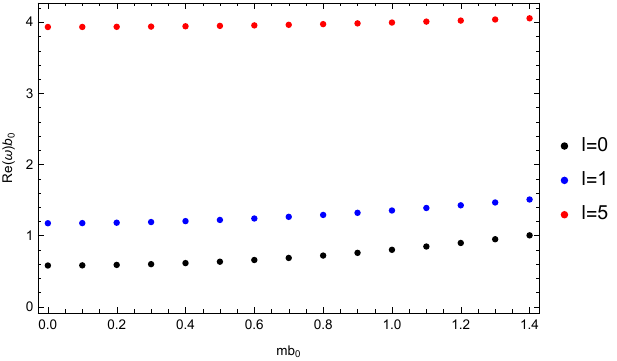}
\includegraphics[width=0.4\textwidth]{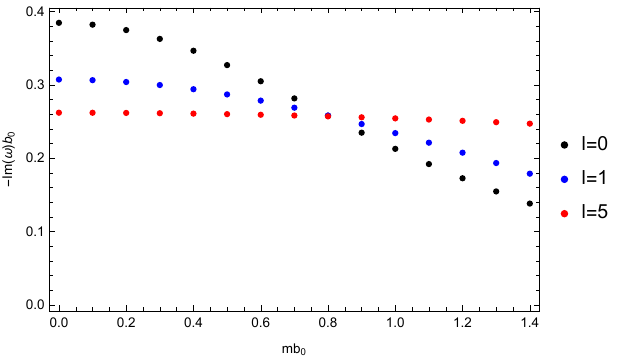}\\
\end{center}
\caption{The behaviour of $Re(\omega)b_0$ and $- Im(\omega)b_0$ for a massive scalar field as a function of $m b_0$ for the wormhole spacetime described by Eq. (\ref{wormhole}), with $M/b_0=0.25$, $n=0$ (fundamental frequency), and $\ell = 0, 1, 5$ using the CFM.}
\label{omega3}
\end{figure}

\begin{figure}[h]
\begin{center}
\includegraphics[width=0.4\textwidth]{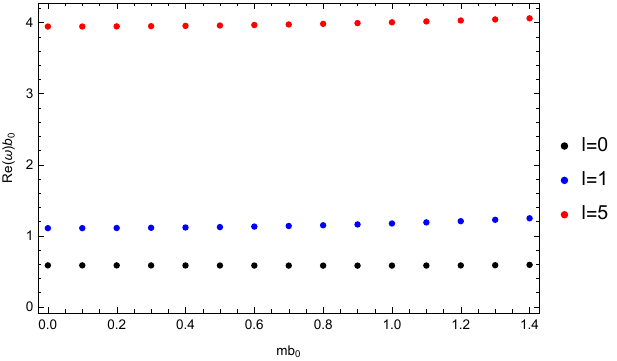}
\includegraphics[width=0.4\textwidth]{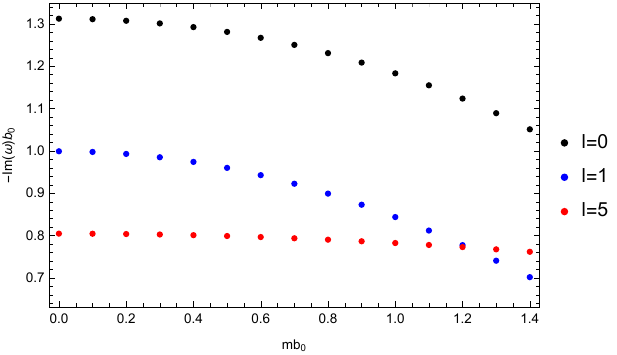}\\
\end{center}
\caption{The behaviour of $Re(\omega)b_0$ and $- Im(\omega)b_0$ for a massive scalar field as a function of $m b_0$ for the wormhole spacetime described by Eq. (\ref{wormhole}), with $M/b_0=0.25$, $n=1$ (first overtone), and $\ell = 0, 1, 5$ using the CFM.}
\label{omega4}
\end{figure}

The continued fraction must be truncated at some large index $N$ and the QNFs are obtained solving this equation numerically.  In the following figures we show the behaviour of $Re(\omega)b_0$ and $- Im(\omega)b_0$ for massive scalar field as a function of $m b_0$ using the CFM, for the fundamental mode, see Figs. \ref{omega_1}, \ref{omega1} and \ref{omega3}, and the first overtone, see Figs. \ref{omega_4}, \ref{omega2} and \ref{omega4}.

For the fundamental QNFs ($n=0$), we can observe an inverted behaviour of $Im(\omega)b_0$, that is, when $m b_0>m_c b_0$, $Im(\omega)b_0$ increases with $\ell$,  representing an anomalous decay rate of the QNMs. Conversely, for $mb_0<m_c b_0$, $Im(\omega)b_0$ decreases with increasing $\ell$ for $n \geq \ell$. The critical value $m_c b_0$ rises with an increase in the parameter $M/b_0$, and the frequency of the oscillation increases with both the angular number and the parameter $mb_0$. The frequencies all have a negative imaginary part, which means that the propagation of massive scalar fields is stable in this background. Some numerical values of the QNFs obtained with CFM are in appendix \ref{NV}. Moreover, in appendix \ref{AFC} we compare the fundamental QNFs and the first overtone using the CFM and the WKB approximation, and we see a good agreement for high values of $\ell$ as expected, because it is known that the WKB method only yields accurate results for high values of $\ell$ and for $\ell > n$.

\subsection{Bound States}

In this section we utilize the CFM to compute the bound states of massive scalar fields around the wormhole geometries defined in (\ref{wormhole}). These bound states are made feasible by the effective potential, which permits the existence of potential wells for $M \neq 0$, see Fig. (\ref{worm}). It is worth to mention that the boundary conditions are different to that of QNMs. In this case we must considering evanescent waves at spatial infinity; therefore, the radial solution that satisfied  the boundary condition at infinity is given by
\begin{equation}
u \sim  e^{-\Omega r} r^{M \omega^2/\Omega} \,\,\,    \text{as} \,\,\, r \rightarrow \infty\,,
\end{equation}
where $\Omega = \sqrt{m^2-\omega^2}$. Generally, in a quantum system with a potential well, there are only a few bound states, the number of which depends on the potential parameters. In several cases, there exists only a single bound state. Thus, we just consider the first bound state which is described by the symmetric solution. Therefore, we consider the following ansatz for the solution to the radial equation (\ref{nueva}) which satisfies the desired boundaries conditions
\begin{equation}
u(r) = e^{- \Omega r} r^{ M \omega^2/\Omega} \sum_{n=0}^{\infty} a_n \left(
\frac{r-b_0}{r} \right)^n \,.
\end{equation}
Then, substituting this expression in the radial equation, the following seven-term recurrence relation is satisfied by the expansion coefficients
\begin{eqnarray} \label{rec}
\notag && c_0(0)a_1+c_1(0)a_0 =0\,, \\
\notag && c_0(1)a_2+c_1(1)a_1+c_2(1)a_0 =0\,, \\
\notag && c_0(2)a_3+c_1(2)a_2+c_2(2)a_1+c_3(2)a_0 =0 \,, \\
\notag && c_0(3)a_4+c_1(3)a_3+c_2(3)a_2+c_3(3)a_1 +c_4(3)a_0 =0 \,, \\
\notag && c_0(4)a_5+c_1(4)a_4+c_2(4)a_3+c_3(4)a_2+c_4(4) a_1+ c_5(4)a_0 =0  \\
&& c_0(n)a_{n+1}+c_1(n)a_{n}+c_2(n)a_{n-1}+c_3(n)a_{n-2}+c_4(n)a_{n-3}+c_5(n)a_{n-4}+c_6(n)a_{n-5} = 0 \,, n \geq 5\,,
\end{eqnarray}
where
\begin{eqnarray}
   \notag c_0(n) &=& (n+1) (2 n+1) \left(b_0-2 M\right) \left(m^2-\omega ^2\right) \,,
\end{eqnarray}
\begin{eqnarray}
   \notag c_1(n) &=& b_0 \Bigg(\omega ^2 \left(\ell^2+\ell-4 M n \sqrt{m^2-\omega ^2}-M \sqrt{m^2-\omega ^2}+9 n^2+1\right)-m^2 \bigg(\ell^2+\ell-8 M n \sqrt{m^2-\omega ^2}-2 M \sqrt{m^2-\omega ^2}  \\
\notag &&  +9 n^2+1\bigg)\Bigg)-b_0^2 \left(m^2-\omega ^2\right) \left((4 n+1) \sqrt{(m-\omega ) (m+\omega )}-2 m^2 M\right)+b_0^3 \left(-\left(m^2-\omega ^2\right)^2\right)+2 M \bigg((\ell^2+\ell  \\
\notag && +11 n^2+1) \left(m^2-\omega ^2\right)-4 M n \omega ^2 \sqrt{m^2-\omega ^2}-M \omega ^2 \sqrt{m^2-\omega ^2}\bigg)  \,,
\end{eqnarray}
\begin{eqnarray}
    \notag c_2(n) &=& 2 b_0 \Bigg(m^2 \left(\ell^2+\ell+2 n \left(-7 M \sqrt{m^2-\omega ^2}+4 n-6\right)+9 M \sqrt{m^2-\omega ^2}+4 M^2 \omega ^2+6\right)-\omega ^2 \bigg(\ell^2+\ell+ \\
\notag &&  n \left(-7 M \sqrt{m^2-\omega ^2}+8 n-12\right)+5 M \sqrt{m^2-\omega ^2}+3 M^2 \omega ^2+6\bigg)\Bigg)+b_0^2 \left(m^2-\omega ^2\right) \bigg((10 n-7)  \\
\notag && \sqrt{(m-\omega ) (m+\omega )}-6 m^2 M\bigg)+2 b_0^3 \left(m^2-\omega ^2\right)^2+M \bigg(-(6 \ell (\ell+1)+25 n (2 n-3)+37) \left(m^2-\omega ^2\right) \\
\notag  && +36 M n \omega ^2 \sqrt{m^2-\omega ^2}-20 M \omega ^2 \sqrt{m^2-\omega ^2}-4 M^2 \omega ^4\bigg) \,,
\end{eqnarray}
\begin{eqnarray}
    \notag c_3(n) &=& b_0 \Bigg(\omega ^2 \left(\ell^2+\ell+2 n \left(-9 M \sqrt{m^2-\omega ^2}+7 n-21\right)+15 M \left(2 \sqrt{m^2-\omega ^2}+M \omega ^2\right)+34\right)-m^2 \bigg(\ell^2+\ell \\
\notag && -6 n \left(6 M \sqrt{m^2-\omega ^2}+7\right)+55 M \sqrt{m^2-\omega ^2}+20 M^2 \omega ^2+14 n^2+34\bigg)\Bigg)+b_0^2 \left(m^2-\omega ^2\right) \bigg(6 m^2 M+(13-8 n)  \\
\notag &&  \sqrt{(m-\omega ) (m+\omega )}\bigg)+b_0^3 \left(-\left(m^2-\omega ^2\right)^2\right)+M \bigg(3 (2 \ell (\ell+1)+20 (n-3) n+49) \left(m^2-\omega ^2\right)- \\
\notag && 64 M n \omega ^2 \sqrt{m^2-\omega ^2}+87 M \omega ^2 \sqrt{m^2-\omega ^2}+14 M^2 \omega ^4\bigg) \,,
\end{eqnarray}
\begin{eqnarray}
  \notag c_4(n) &=& b_0 \Bigg(m^2 \left(n \left(-20 M \sqrt{m^2-\omega ^2}+6 n-27\right)+16 M \left(3 \sqrt{m^2-\omega ^2}+M \omega ^2\right)+31\right)-\omega ^2 \bigg(n (-10 M \sqrt{m^2-\omega ^2}+  \\
\notag && 6 n-27)+26 M \sqrt{m^2-\omega ^2}+12 M^2 \omega ^2+31\bigg)\Bigg)+b_0^2 \left(m^2-\omega ^2\right) \left((2 n-5) \sqrt{(m-\omega ) (m+\omega )}-2 m^2 M\right)+ \\
\notag && M \left(-(2 \ell (\ell+1)+20 n (2 n-9)+209) \left(m^2-\omega ^2\right)+56 M n \omega ^2 \sqrt{m^2-\omega ^2}-121 M \omega ^2 \sqrt{m^2-\omega ^2}-18 M^2 \omega ^4\right) \,,
\end{eqnarray}
\begin{eqnarray}
   \notag c_5(n) &=& b_0 \Bigg(\omega ^2 \left(n \left(-2 M \sqrt{m^2-\omega ^2}+n-6\right)+7 M \sqrt{m^2-\omega ^2}+3 M^2 \omega ^2+9\right)-m^2 \bigg(n \left(-4 M \sqrt{m^2-\omega ^2}+n-6\right)  \\
\notag && +13 M \sqrt{m^2-\omega ^2}+4 M^2 \omega ^2+9\bigg)\Bigg)+M \bigg(-24 M n \omega ^2 \sqrt{m^2-\omega ^2}+71 M \omega ^2 \sqrt{m^2-\omega ^2}+(14 (n-6) n+127)  \\
\notag && \left(m^2-\omega ^2\right)+10 M^2 \omega ^4\bigg) \,,
\end{eqnarray}
\begin{eqnarray}
  \notag c_6(n) &=& M \left(\omega ^2 \left(n \left(4 M \sqrt{m^2-\omega ^2}+2 n-15\right)-15 M \sqrt{m^2-\omega ^2}-2 M^2 \omega ^2+28\right)-m^2 (n-4) (2 n-7)\right) \,.
\end{eqnarray}

Then, after applying Gaussian elimination four times, the original seven-term recurrence relations (\ref{rec}) can be simplified to a three-term relation, taking on the following concise form
\begin{eqnarray}
\notag \alpha_0 a_1 + \beta_0 a_0 &=& 0  \,, \\
\notag \alpha_n a_{n+1} + \beta_n a_n + \gamma_n a_{n-1} &=&  0 \,, \,\,\,\, n = 1, 2, ...
\end{eqnarray}
We skip providing the explicit expressions for $\alpha_n$, $\beta_n$ and $\gamma_n$ of the final three-term recurrence relations.

Now, according to Leaver \cite{Leaver:1985ax, Leaver:1990zz}, the recursion coefficients must satisfy the following continued fraction relation for the convergence of the series
\begin{equation}
\beta_0 - \frac{\alpha_0 \gamma_1}{\beta_1-}\frac{\alpha_1 \gamma_2}{\beta_2-} \cdots\frac{\alpha_n \gamma_{n+1}}{\beta_{n+1}-} \cdots = 0\,,
\end{equation}
and the continued fraction must be truncated at some large index $N$. The bound states frequencies are obtained solving this equation numerically.
In Tables \ref{TableI} and \ref{NVMT3}, we show some values of the lowest bound states frequencies $\omega b_0$ for scalar perturbations using the CFM. We observe that the oscillation frequency increases when the mass parameter $mb_0$ increases and when the angular number $\ell$ increases.
\begin{table}[h]
\centering
\caption{Frequencies ($\omega b_0$) of the lowest bound states for scalar perturbations in the wormhole spacetime described by Eq. (\ref{wormhole}), with $M/b_0=0.1$, $\ell =0, 1$, and several values of the scalar field mass using the CFM.}
\label{TableI}
\begin{tabular}{|c|c|c|c|c|}
\hline
$\ell$ &  $m b_0=0.4$ & $m b_0=0.8$ & $m b_0=1.2$ & $m b_0 = 1.6$ \\
%\colrule
\hline
0  & 0.39967670 & 0.79727718 &  1.18979435  &  1.57334204 \\
1  & 0.39991988  & 0.79935626 & 1.19781061 &  1.59474718 \\
  \hline
\hline
$\ell$ &  $m b_0=2$ & $m b_0=2.4$ & $m b_0=2.8$ & $m b_0 = 3.2$  \\
%\colrule
\hline
0  & 1.94709369 & 2.31390013 & 2.67665885 & 3.03714485 \\
1  & 1.98954025 & 2.38132146 & 2.76858541 & 3.14873064  \\
\hline
\end{tabular}
\end{table}

\begin{table}[h]
\centering
\caption{Frequencies ($\omega b_0$) of the lowest bound states for scalar perturbations in the wormhole spacetime described by Eq. (\ref{wormhole}), with $M/b_0 =0.25$, $\ell =0, 1$, and several values of the scalar field mass using the CFM.}
\label{NVMT3}
\begin{tabular}{|c|c|c|c|c|}
\hline
$\ell$ &  $m b_0=0.4$ & $m b_0=0.8$ & $m b_0=1.2$ & $m b_0 = 1.6$ \\
%\colrule
\hline
0  & 0.39948020 & 0.77659459 &  1.10626752  &  1.40450402 \\
1  & 0.39949541  & 0.79583960 & 1.18498187 &  1.55693415 \\
  \hline
\hline
$\ell$ &  $m b_0=2$ & $m b_0=2.4$ & $m b_0=2.8$ & $m b_0 = 3.2$  \\
%\colrule
\hline
0  & 1.69213119 & 1.97599183 & 2.25838693 & 2.54020128 \\
1  & 1.87946685   & 2.16141735 & 2.43284661 & 2.70240221  \\
\hline
\end{tabular}
\end{table}
%\newpage

\section{f(R) gravity wormhole}
\label{sec5}

 In the framework of $f(R)$ modified gravity, a generalization of the Ellis drainhole solution was found in \cite{Karakasis:2021tqx} by considering a phantom scalar field with a self-interacting potential. The solution is given by
\begin{equation}
\Phi(r) = 0 \,,\,\,
b(r) = 2r - \mu^2 r^3 \,,\,\,
\phi(r) = \sqrt{6} \frac{\sqrt{A}}{r} \,,\,\,
V(\phi) = \frac{1}{2} \mu^2 \phi^2\,.\,\,
\end{equation}
The condition $\frac{b(r)}{r} \leq 1$ reduces to $\mu^2 r^2 \geq 1$, which holds for all $r$ within the interval $[\frac{1}{\mu}, + \infty)$. Also, the condition $\frac{b(r)}{r}=1$ yields the throat location at $r_{th} = \frac{1}{\mu}$, where $\mu$ can be interpreted as the mass of the phantom scalar field. Furthermore, the flare-out condition $b'(r)< \frac{b(r)}{r}$ results in $\mu^2 r^2 >0$, which is satisfied for every $r>0$ and $\mu \neq 0$. Also, it can be easily verified that $b'(r_{th})=-1$, ensuring that the relation $b'(r_{th}) \leq 1$ always holds true.

The fundamental characteristics of the solution highlight the impact of the phantom scalar field on both the scalar curvature and the throat's size. As the strength of the scalar field intensifies, the resulting wormholes exhibit larger throat radii, consequently diminishing the curvature in the vicinity of the throat. We direct the reader to Ref. \cite{Karakasis:2021tqx} for a comprehensive review of the solution.

\subsection{Analytical QNFs of f(R) gravity wormhole}

\begin{figure}[h!]
\begin{center}
\includegraphics[width=0.5\textwidth]{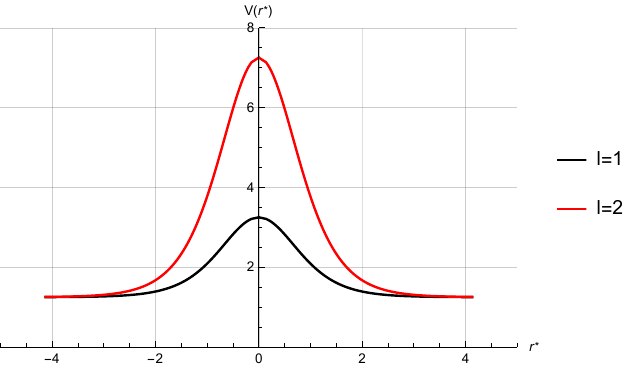}\\
\end{center}
\caption{Effective potential for $\mu=1$, $m=0.5$ and $\ell = 1, 2$.
}
\label{worm2}
\end{figure}

 In this section we study the propagation of a probe scalar field in the $f(R)$ wormhole geometry, with a particular focus on studying its stability through the QNFs. In this case the Klein-Gordon equation, considering the following separation of variables
\begin{equation}\label{separable}
\Psi(t,r,\theta,\varphi) = e^{-i \omega t} \: \frac{u(r)}{r} \:Y_{\ell} (\Omega)\,,
\end{equation}can be written as a one-dimensional Schr{\"o}dinger-like equation
\begin{equation}
\frac{\mathrm{d}^2 u}{\mathrm{d}r^{\ast}{}^2} + [ \omega^2 - V(r^{\ast}) ] u = 0\,,
\label{sequation}
\end{equation}
where $\omega$ is the unknown frequency, while $Y_{\ell}(\Omega)$ are the spherical harmonics, and $r^{\ast}$ is the well-known tortoise coordinate, i.e.,
\begin{equation}
r^{\ast}  \equiv  \int \frac{\mathrm{d}r}{e^{\Phi} \sqrt{1 -\frac{b(r)}{r}}}\,.
\end{equation}
The effective potential governing scalar perturbations for the $f(R)$ gravity wormhole reads
\begin{equation}
V(r) = \frac{\ell (\ell+1)}{r^2} + \mu^2 + m^2\,,
\end{equation}
which is plotted in Fig. \ref{worm2}.  We observe that the effective potential is symmetric about the throat of the wormhole and takes the form of a barrier for $\ell \neq 0$, with the barrier magnitude increasing with $\ell$. However, it does not allow wells, consequently precluding the support for bound states.

It is worth noticing that when $\ell=0$ the effective potential is constant. Consequently, the solution of the radial function is given by $u = C_1 e^{i \sqrt{\omega^2-\mu^2-m^2} r^*} + C_2 e^{- i \sqrt{\omega^2 - \mu^2 - m^2} r^*}$. Imposing the condition of purely outgoing waves at both infinities leads to the conclusion that $C_1 = C_2 = 0$, resulting in $u=0$ for this particular case.
%\newline

Now we consider the case $\ell \neq 0$. The radial equation can be written as
\begin{equation}
(1-\tilde{r}^2) \frac{d^2}{d\tilde{r}^2} u(\tilde{r}) -  \tilde{r}  \frac{d}{d \tilde{r}} u(\tilde{r}) + \left( \frac{\ell (\ell +1)}{\tilde{r}^2} -\tilde{\omega}^2 + 1 + \tilde{m}^2  \right)u(\tilde{r}) =0\,,
\end{equation}
where we have defined a dimensionless coordinate $\tilde{r}= \mu r$, $\tilde{m} = m / \mu$ and $\tilde{\omega} = \omega / \mu$. The radial equation can be analytically solved by expressing it in terms of hypergeometric functions
\begin{eqnarray}
\notag u(\tilde{r}) &=& C_1 \tilde{r}^{\frac{1}{2} (1- \sqrt{1-4 \ell (\ell+1)})} \,_2F_1 \Bigg( \frac{1}{4} - \frac{1}{4} \sqrt{1-4 \ell (\ell+1)} -\frac{\sqrt{1 + \tilde{m}^2 -\tilde{\omega}^2}}{2}, \frac{1}{4} - \frac{1}{4} \sqrt{1-4 \ell (\ell+1)} + \frac{\sqrt{1 + \tilde{m}^2 -\tilde{\omega}^2}}{2},  \\
\notag && 1- \frac{1}{2}\sqrt{1-4 \ell (\ell+1)}, \tilde{r}^2 \Bigg) + C_2 r^{\frac{1}{2} (1+ \sqrt{1-4 \ell (\ell+1)})} \,_2F_1 \Bigg( \frac{1}{4} + \frac{1}{4} \sqrt{1-4 \ell (\ell+1)} -\frac{\sqrt{1 + \tilde{m}^2 -\tilde{\omega}^2}}{2},  \\
&& \frac{1}{4} + \frac{1}{4} \sqrt{1-4 \ell (\ell+1)} + \frac{\sqrt{1 + \tilde{m}^2 -\tilde{\omega}^2}}{2}, 1+ \frac{1}{2}\sqrt{1-4 \ell (\ell+1)}, \tilde{r}^2 \Bigg) \,,
\end{eqnarray}
where $\, _2F_1$ is the Gaussian Hypergeometric function.

Since the potential is symmetric about the throat of the wormhole, which is located at $r_0^{\ast}=0$, the solutions will be symmetric or anti-symmetric. Therefore, we impose the following boundary conditions at the throat, $\frac{du}{d r^{\ast}}\left|_{r_0^{\ast}}=0\right.$ for the symmetric solutions and $u(r_0^{\ast}) = 0$ for the anti-symmetric solutions. These boundary conditions yields the even and odd overtones, respectively. For the symmetric solutions the condition at the throat
$\frac{du}{dr^{\ast}}\left|_{r_0^{\ast}}\right. = \frac{du}{d\tilde{r}} \sqrt{1- \frac{b(\tilde{r})}{\tilde{r}}}\left|_{\tilde{r}= 1} \right. = 0$
yields
\begin{equation}
\notag C_1 = C_2 \rho\frac{\Gamma[1+\alpha_{+}] \Gamma[1+\alpha_{-}] \Gamma \left[  1 + \frac{1}{2} \sqrt{1- 4 \ell (\ell+1)} \right]}{ \Gamma [1+\gamma_{-}]  \Gamma [1+\gamma_{+}] \Gamma \left[  1 - \frac{1}{2} \sqrt{1- 4 \ell (\ell+1)} \right]}  \,,
\end{equation}
where
\begin{equation}
\notag  \rho =   \frac{ -2 (2 -\sqrt{1- 4\ell (\ell+1)}) \tilde{m}^2 - 3 + 4 \tilde{\omega}^2 + \sqrt{1- 4 \ell(\ell+1)} ((3+ 2 \ell (\ell+1)) - 2\tilde{\omega}^2)}{ \left( 2 \tilde{m}^2 + (1+ 2\ell (\ell+1) + \sqrt{1- 4\ell (\ell+1)}) - 2 \tilde{\omega}^2 \right) (2 - \sqrt{1- 4\ell (\ell+1)})}\,,
\end{equation}
and the asymptotic behaviour of the solutions is
\begin{eqnarray}
\notag u(\tilde{r}) \sim &&  C_2 \tilde{r}^{- i\sqrt{\tilde{\omega}^2 -1 -\tilde{m}^2}}  \Gamma \left[ -\sqrt{1+\tilde{m}^2-\tilde{\omega}^2} \right]   \Gamma \left[  1 + \frac{1}{2} \sqrt{1- 4 \ell (\ell+1)} \right] \Bigg( (-1)^{-\alpha_{+}} \rho (\alpha_{-} / \gamma_{-} )\frac{\Gamma[1+\alpha_{+}]}{  \Gamma [\beta_{-}]   \Gamma [1+\gamma_{+}]}  \\
\notag &&  +  \frac{ (-1)^{-\gamma_{+}} }{ \Gamma[ \delta_{-}]} \Bigg) \frac{1}{\Gamma [\gamma_{-}]} + C_2 \tilde{r}^{i\sqrt{\tilde{\omega}^2 - 1 -\tilde{m}^2}}  \Gamma \left[ \sqrt{1+\tilde{m}^2-\tilde{\omega}^2} \right] \Gamma \left[  1 + \frac{1}{2} \sqrt{1- 4 \ell (\ell+1)} \right]  \\
&&  \Bigg( (-1)^{-\alpha_{-}} \rho  (\alpha_{+}/\gamma_{+})  \frac{ \Gamma[1+\alpha_{-}] }{ \Gamma[\beta_{+}] \Gamma [1+\gamma_{-}]  } +  \frac{ (-1)^{-\gamma_{-}}}{  \Gamma[\delta_{+}]}\Bigg) \frac{1}{ \Gamma [\gamma_{+}]}\,,
\end{eqnarray}
where we have defined
\begin{eqnarray}
\notag \alpha_{\pm} &=& \frac{1}{4} - \frac{\sqrt{1- 4 \ell (\ell+1)}}{4} \pm \frac{\sqrt{1+\tilde{m}^2-\tilde{\omega}^2}}{2} \,, \,\,\,\, \notag \beta_{\pm} = \frac{3}{4} - \frac{\sqrt{1- 4 \ell (\ell+1)}}{4} \pm \frac{\sqrt{1+ \tilde{m}^2-\tilde{\omega}^2}}{2}\,, \\
\notag \gamma_{\pm} &=& \frac{1}{4} + \frac{\sqrt{1- 4 \ell (\ell+1)}}{4} \pm \frac{\sqrt{1+\tilde{m}^2-\tilde{\omega}^2}}{2} \,, \,\,\,\, \notag \delta_{\pm} = \frac{3}{4} + \frac{\sqrt{1- 4 \ell (\ell+1)}}{4} \pm \frac{\sqrt{1+\tilde{m}^2-\tilde{\omega}^2}}{2}  \,.
\end{eqnarray}

Now, in accordance with the boundary condition of purely outgoing waves at spatial infinity, i.e. $u \sim r^{i \sqrt{\tilde{\omega}^2- 1 -\tilde{m}^2}}$, we set
\begin{equation}
\gamma_{-} = \frac{1}{4} + \frac{\sqrt{1- 4 \ell (\ell+1)}}{4} -\frac{\sqrt{1+\tilde{m}^2-\tilde{\omega}^2}}{2} = -n \,\,\,\,\,\,\,\, n=0,1, ...
%\cdots
\end{equation}
Therefore, the QNFs for even overtones yield
\begin{equation} \label{qnf1}
\tilde{\omega} = \frac{1}{\sqrt{2}} \left( \eta + \left(\eta^2 + \frac{1}{4} (-1 + 4 \ell (\ell+1))(1+4 n)^2 \right)^{1/2} \right)^{1/2}- i \frac{\left(-1+4 \ell (\ell+1) \right)^{1/2} (1+ 4 n)}{2 \sqrt{2} \left( \eta + \left(\eta^2 + \frac{1}{4}(-1+4 \ell (\ell+1))(1+4n)^2 \right)^{1/2} \right)^{1/2}} \,,
\end{equation}
where $\eta \equiv \frac{1}{2}(1+ 2 \ell (\ell+1)-4 n (1+2n))+\tilde{m}^2$ \,.
\newline

On the other hand, the boundary condition at the throat $u(1) = 0$ for the antisymmetric solutions leads to the expression
\begin{equation}
\notag C_1 = -C_2 \frac{\Gamma[\beta_{+}] \Gamma[\beta_{-}] \Gamma \left[  1 + \frac{1}{2} \sqrt{1- 4 \ell (\ell+1)} \right]}{ \Gamma [\gamma_{+}]  \Gamma [\gamma_{-}] \Gamma \left[  1 - \frac{1}{2} \sqrt{1- 4 \ell (\ell+1)} \right]}  \,.
\end{equation}
Thus, the asymptotic behaviour of the solutions is
\begin{eqnarray}
\notag u(\tilde{r}) \sim &&  C_2 \tilde{r}^{- i\sqrt{\tilde{\omega}^2 -1 -\tilde{m}^2}}  \Gamma \left[ -\sqrt{1+\tilde{m}^2-\tilde{\omega}^2} \right]   \Gamma \left[  1 + \frac{1}{2} \sqrt{1- 4 \ell (\ell+1)} \right] \Bigg( -(-1)^{-\alpha_{+}} \frac{\Gamma[\beta_{+}]}{  \Gamma [\alpha_{-}]   \Gamma [\delta_{+}]}  \\
\notag &&  +  \frac{ (-1)^{-\gamma_{+}} }{ \Gamma[ \gamma_{-}]} \Bigg) \frac{1}{\Gamma [\delta_{-}]} + C_2 \tilde{r}^{i\sqrt{\tilde{\omega}^2 - 1 -\tilde{m}^2}}  \Gamma \left[ \sqrt{1+\tilde{m}^2-\tilde{\omega}^2} \right] \Gamma \left[  1 + \frac{1}{2} \sqrt{1- 4 \ell (\ell+1)} \right]  \\
&&  \Bigg( -(-1)^{-\alpha_{-}} \frac{ \Gamma[\beta_{-}] }{ \Gamma[\alpha_{+}] \Gamma [\delta_{-}]  } +  \frac{ (-1)^{-\gamma_{-}}}{  \Gamma[\gamma_{+}]}\Bigg) \frac{1}{ \Gamma [\delta_{+}]}\,.
\end{eqnarray}
Now, in accordance with the boundary condition of purely outgoing waves at spatial infinity, i.e. $u \sim r^{i \sqrt{\tilde{\omega}^2- 1 -\tilde{m}^2}}$, we set
\begin{equation}
\delta_{-} = \frac{3}{4} + \frac{\sqrt{1- 4 \ell (\ell+1)}}{4} -\frac{\sqrt{1+\tilde{m}^2-\tilde{\omega}^2}}{2} = -n \,\,\,\,\,\,\,\, n=0,1, ...
\end{equation}
and the QNFs for odd overtones yield
\begin{equation} \label{qnf2}
\tilde{\omega} = \frac{1}{\sqrt{2}} \left( \eta' + \left(\eta'^2 + \frac{1}{4} (-1 + 4 \ell (\ell+1))(3+4 n)^2 \right)^{1/2} \right)^{1/2}- i \frac{\left(-1+4 \ell (\ell+1) \right)^{1/2} (3+ 4 n)}{2 \sqrt{2} \left( \eta' + \left(\eta'^2 + \frac{1}{4}(-1+4 \ell (\ell+1))(3+4n)^2 \right)^{1/2} \right)^{1/2}} \,,
\end{equation}
where $\eta' \equiv \frac{1}{2}(-3+ 2 \ell (\ell+1)-4 n (3+2n))+\tilde{m}^2$.
\newline

\begin{figure}[h!]
\begin{center}
\includegraphics[width=0.4\textwidth]{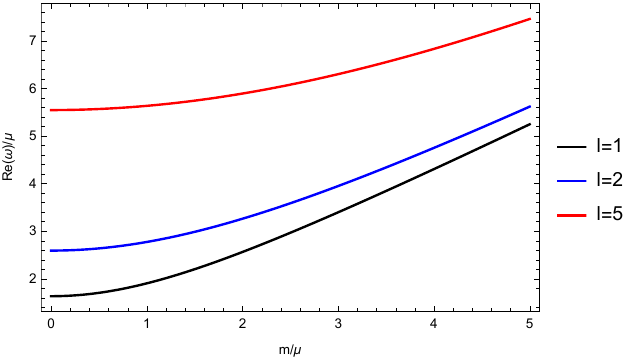}
\includegraphics[width=0.4\textwidth]{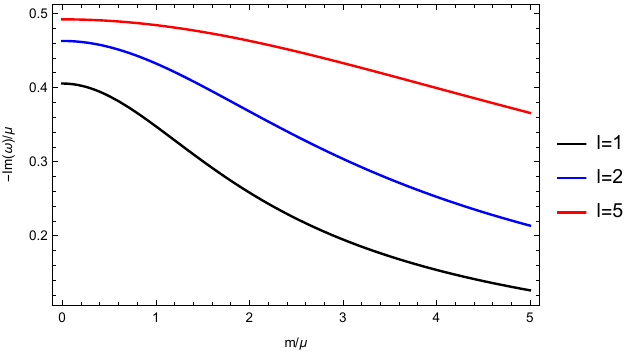}
\end{center}
\caption{The behaviour of $Re(\omega)/\mu$ (left panel), and $-Im(\omega)/\mu$ (right panel) as a function of $m / \mu$ for different values of $\ell$ and $n=0$.}
\label{frr1}
\end{figure}

\begin{figure}[h!]
\begin{center}
\includegraphics[width=0.4\textwidth]{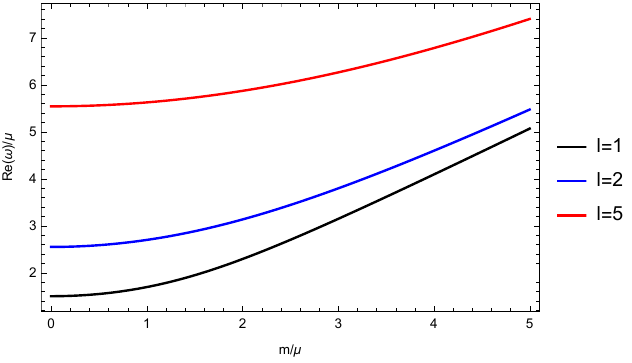}
\includegraphics[width=0.4\textwidth]{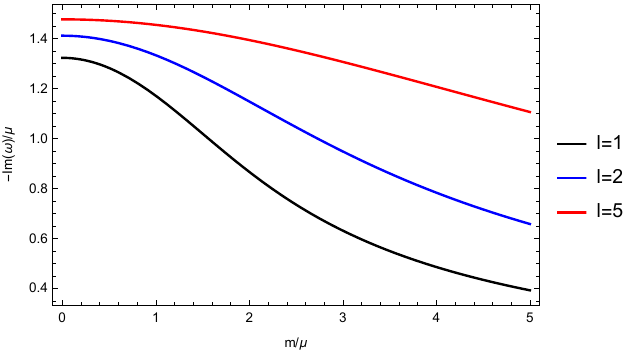}
\end{center}
\caption{The behaviour of $Re(\omega)/\mu$ (left panel), and $-Im(\omega)/\mu$ (right panel) as a function of $m / \mu$ for different values of $\ell$ and $n=1$.}
\label{frr2}
\end{figure}

Finally, the QNFs given by Eq. (\ref{qnf1}) for even overtones and Eq. (\ref{qnf2}) for odd overtones, can be succinctly expressed in a unified formula as
\begin{equation} \label{analytical}
\tilde{\omega} = \frac{1}{\sqrt{2}} \left( \eta'' + \left(\eta''^2 + \frac{1}{4} (-1 + 4 \ell (\ell+1))(1+2 n)^2 \right)^{1/2} \right)^{1/2}- i \frac{\left(-1+4 \ell (\ell+1) \right)^{1/2} (1+ 2 n)}{2 \sqrt{2} \left( \eta'' + \left(\eta''^2 + \frac{1}{4}(-1+4 \ell (\ell+1))(1+2n)^2 \right)^{1/2} \right)^{1/2}} \,,
\end{equation}
where $\eta'' \equiv \frac{1}{2}(1+ 2 \ell (\ell+1)-2 n (1+n))+\tilde{m}^2$ and $n=0,1, \dots$. In Figs. \ref{frr1} and \ref{frr2} we plot the behaviour of $Re(\tilde{\omega})$, and $-Im(\tilde{\omega})$ as a function of $\tilde{m}$ by using the analytical expression (\ref{analytical}) for $n=0$ and $n=1$, respectively.
%, and $n=1$, respectively.
We can observe that the modes with the lowest angular number $\ell$ exhibit the longest lifetimes, both for the fundamental and the first overtone, and the decay rate decreases with $\tilde{m}$. Additionally, the frequency of oscillation increases with higher values of the parameter $\tilde{m}$ and the angular momentum $\ell$.

All the frequencies exhibit a negative imaginary part, indicating the stability of the propagation of massive scalar fields in this background. Defining $L=\sqrt{\ell (\ell+1)}$ and considering high values of $L$, this expression can be approximated to
\begin{equation} \label{analitica}
\tilde{\omega} \approx L + \frac{3+4 \tilde{m}^2}{8L} - i \left(  \frac{1}{2}+n - \frac{(1+\tilde{m}^2)(1+2n)}{4 L^2}  \right) \,.
\end{equation}
As we will see in the next subsection this result match with the QNFs obtained via the third order WKB method.

\subsection{WKB Method}

 In this section, we employ the Wentzel-Kramers-Brillouin (WKB) approximation method, first introduced by Mashhoon \cite{Mashhoon} and later developed by Schutz and Iyer \cite{Schutz:1985km}, to obtain the QNFs. This approach allows us to gain insights into the behaviour of QNFs in the eikonal limit as $\ell \rightarrow \infty$. Our aim is to validate and complement the analytical results obtained in the previous section using an alternative method. Iyer and Will computed the third order correction \cite{Iyer:1986np}, which was subsequently extended to the sixth order \cite{Konoplya:2003ii}, and recently up to the 13th order \cite{Matyjasek:2017psv}, see also \cite{Konoplya:2019hlu}.

\begin{figure}[h!]
\begin{center}
\includegraphics[width=0.4\textwidth]{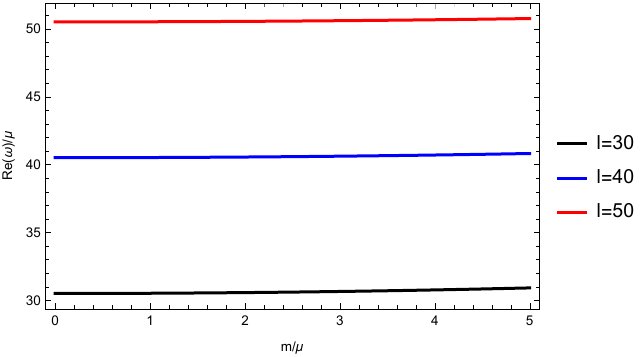}
\includegraphics[width=0.4\textwidth]{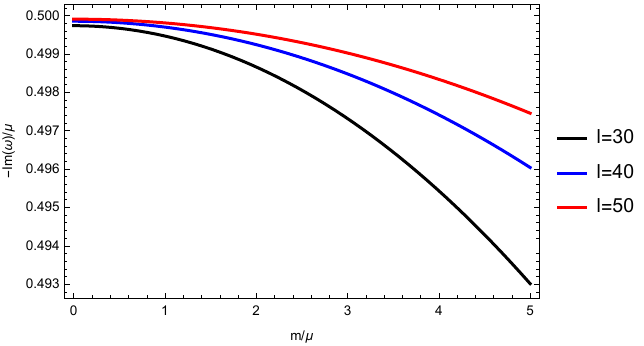}
\end{center}
\caption{The behaviour of $Re(\omega)/\mu$ (left panel), and $-Im(\omega)/\mu$ (right panel) as a function of $m / \mu$ for different values of $\ell$ using the six order WKB method and $n=0$.}
\label{fr1}
\end{figure}

\begin{figure}[h!]
\begin{center}
\includegraphics[width=0.4\textwidth]{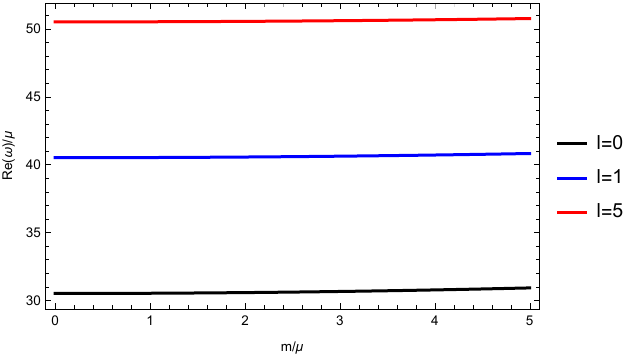}
\includegraphics[width=0.4\textwidth]{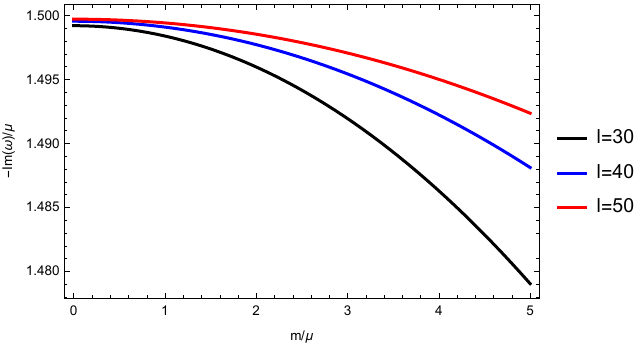}
\end{center}
\caption{The behaviour of $Re(\omega)/\mu$ (left panel), and $-Im(\omega)/\mu$ (right panel) as a function of $m / \mu$ for different values of $\ell$ using the six order WKB method and $n=1$.}
\label{fr2}
\end{figure}

This method has been used to determine the QNFs for asymptotically flat and asymptotically de Sitter black holes. This is due to the WKB method can be used for effective potentials which have the form of potential barriers that approach to a constant value at the horizon and spatial infinity \cite{Konoplya:2011qq}. The QNMs are determined by the behaviour of the effective potential near its maximum value $V(r^*_{max})$. The Taylor series expansion of the potential around its maximum is given by
\begin{equation}
V(r^*)= V(r^*_{max})+ \sum_{i=2}^{\infty} \frac{V^{(i)}}{i!} (r^*-r^*_{max})^{i} \,,
\end{equation}
where
\begin{equation}
V^{(i)}= \frac{d^{i}}{d r^{*i}}V(r^*)|_{r^*=r^*_{max}}\,,
\end{equation}
corresponds to the $i$-th derivative of the potential with respect to $r^*$ evaluated at the position of the maximum of the potential $r^*_{max}$. Using the WKB approximation up to third order beyond the eikonal limit, the QNFs are given by the following expression \cite{Hatsuda:2019eoj}
\begin{eqnarray}
\omega^2 &=& V(r^*_{max})  -2 i U \,,
\end{eqnarray}
where
\begin{eqnarray}
\notag U &=&  N\sqrt{-V^{(2)}/2}+\frac{i}{64} \left( -\frac{1}{9}\frac{V^{(3)2}}{V^{(2)2}} (7+60N^2)+\frac{V^{(4)}}{V^{(2)}}(1+4 N^2) \right) +\frac{N}{2^{3/2} 288} \Bigg( \frac{5}{24} \frac{V^{(3)4}}{(-V^{(2)})^{9/2}} (77+188N^2)  \\
\notag && +\frac{3}{4} \frac{V^{(3)2} V^{(4)}}{(-V^{(2)})^{7/2}}(51+100N^2) +\frac{1}{8} \frac{V^{(4)2}}{(-V^{(2)})^{5/2}}(67+68 N^2)+\frac{V^{(3)}V^{(5)}}{(-V^{(2)})^{5/2}}(19+28N^2)+\frac{V^{(6)}}{(-V^{(2)})^{3/2}} (5+4N^2)  \Bigg)\,,
\end{eqnarray}
and $N=n+1/2$, with $n=0,1,2,\dots$, is the overtone number. Now, defining $L^2= \ell (\ell+1)$, we find that for large values of $L$, the QNFs are approximately given by
\begin{equation}
\tilde{\omega} \approx \omega_1 L + \omega_0 + \omega_{-1} L^{-1} + \omega_2 L^{-2} + \mathcal{O}(L^{-3})\,.
\end{equation}
The maximum of the potential is located at the throat $\tilde{r}= 1$ and its value is given by
\begin{equation}
 V(r^*_{0})=  L^2 + 1+ \tilde{m}^2 \,,
\end{equation}
and the derivatives of the potential are
\begin{equation}
V^{(2)}(r_0^{\ast}) = -2 L^2 \,, \,\,\,\,\, V^{(3)}(r_0^{\ast}) = 0  \,, \,\,\,\,\, V^{(4)}(r_0^{\ast}) = 16 L^2 \,, \,\,\,\,\, V^{(5)}(r_0^{\ast}) = 0  \,, \,\,\,\,\, V^{(6)}(r_0^{\ast}) = -272 L^2\,.
\end{equation}
Therefore, the QNFs for large $L$ are approximately given by
\begin{equation}
\tilde{\omega} \approx L + \frac{3+4 \tilde{m}^2}{8L} - i \left(  \frac{1}{2}+n - \frac{(1+\tilde{m}^2)(1+2n)}{4 L^2}  \right) \,.
\end{equation}
This result, as previously mentioned, coincides with the earlier finding, as shown in Eq. (\ref{analitica}). Now, we plot the behaviour of $Re(\omega)/\mu$, and $-Im(\omega/\mu)$ as a function of $m/\mu$ by using the 6th order WKB  method for $n=0$ (see, Fig. \ref{fr1}), and $n=1$ (see, Fig. \ref{fr2}).
%, and $n=1$, respectively.
We can observe that the modes with the lowest angular number $\ell$ exhibit the longest lifetimes, both for the fundamental and the first overtone. Additionally, the frequency of oscillation increases with higer values of the parameter $m/\mu$ and angular momentum $\ell$. Therefore, in this wormhole geometry the anomalous decay of QNFs is avoided.

%\newpage

%%%%%%%%%%%%%%%%%%%%%%%%%%%%%%%%%%%%%%%%%%%%%%%%%%%%%%%%%%%%%%%%%%%%%%%%%

\section{Conclusions}

\label{sec6}

In this work, we explored wormhole geometries with a non-constant redshift function by introducing a gravitational mass $M$, which goes over into the Bronnikov-Ellis wormhole when the gravitational mass parameter vanishes and an $f(R)$ gravity wormhole spacetime as background geometries and investigated the propagation of massive scalar fields.

For the wormholes with a non-constant reshift function, our analysis focused primarily on low values of the angular number $\ell$, and we determined the QNFs using the CFM. These QNFs exhibit an inverted behaviour of $Im(\omega)b_0$: when $m b_0>m_c b_0$, $Im(\omega)b_0$ increases with $\ell$, representing an anomalous decay rate of the QNMs. Conversely, for $mb_0<m_c b_0$, $Im(\omega)b_0$ decreases with increasing $\ell$ for $n \geq \ell$. The critical value $m_c b_0$ rises with an increase in the parameter $M/b_0$, and the frequency of the oscillation increases with both the angular number and the parameter $mb_0$. Additionally, for the geometries with $M\neq 0$ we identified bound states characterized by an oscillation frequency without decay that rises with increasing $mb_0$ and $\ell$. On the other hand, for $M=0$ the effective potential does not forms wells, and bound states are prohibited in this case.

Then, we considered a wormhole spacetime that is an exact solution of $f(R)$ modified gravity with a phantom scalar field with a self-interacting potential. We obtained exact results for the QNFs, which, for large values of $\ell$, coincide with the third-order WKB approximation. Here,  the longest-lived modes are those with the lowest angular number $\ell$, both for the fundamental and the first overtone. Furthermore, the oscillation frequency increases with a rise in the parameter $m/\mu$ and angular momentum $\ell$. Thus, in this wormhole background, the anomalous behaviour observed in the first spacetimes is avoided. Also, the effective potential is symmetric about the throat of the wormhole and it does not allow wells, consequently bound states are not supported in this wormhole spacetime. 

The analysis of the wormholes considered reveals that those with a constant redshift function do not support bound states, in contrast to those with a non-constant redshift function (resulting in a non-zero tidal radial force), which do permit bound states.

%%%%%%%%%%%%%%%%%%%%%%%%%%%%%%%%%%%%%%%%%%%%%%%%%%%%%%%%%%%%%%%%%%%%%%%%%%%%%%

\section*{Acknowlegements}
\noindent
This work is partially supported by ANID Chile through FONDECYT Grant Nº 1220871  (P.A.G., and Y. V.). P. A. G. acknowledges the hospitality of the Universidad de La Serena where part of this work was undertaken.

%%%%%%%%%%%%%%%%%%%%%%%%%%%%%%%%%%%%%%%%%%%%%%%%%%%%%%%%%%%%%%%%%%%%%%%%%%%%%%

%\newpage

\appendix

\section{Numerical values}
\label{NV}

In this appendix we provide some numerical values of the QNFs via CFM for the wormhole spacetime described by Eq. (\ref{wormhole}). In Table \ref{NVBE1} and \ref{NVMT0} we show the QNFs for the fundamental mode for $M/b_0 =0.1$, and $M/b_0 =0.25$, respectively, and in Table \ref{NVBE12} and \ref{NVMT02} we show the QNFs for the first overtone ($n=1$) for $M/b_0 =0.1$, and $M/b_0 =0.25$, respectively.

\begin{table}[h]
\centering
\caption{Fundamental $n=0$ QNFs $\omega b_0$ for scalar perturbations for the Bronnikov-Ellis wormhole for several values of the angular momentum and the mass of the scalar field using the CFM.}
\label{TableIII}
\begin{tabular}{|c|c|c|c|c|}
\hline
$\ell$ &  $m b_0=0$ & $m b_0=0.2$ & $m b_0=0.4$ & $m b_0 = 0.6$ \\
%\colrule
\hline
0  & 0.68136719-0.61775345 i & 0.69762539-0.60335666 i  & 0.74791262-0.56278891 i  & 0.83479189-0.50421780 i  \\
1  & 1.57270839-0.52970088 i  & 1.58410509-0.52589000 i  & 1.61799977-0.51487339 i &  1.67350804-0.49779565 i \\
2   & 2.54665642-0.51266881 i & 2.55419456-0.51115579 i  & 2.57669677-0.50669188 i & 2.61383299-0.49949302 i  \\
3  & 3.53428741-0.50692165 i   & 3.53982819-0.50612818 i & 3.55640289-0.50376936 i & 3.58387030-0.49990838 i  \\
5  & 5.52230920-0.50295823 i & 5.52589995-0.50263141 i & 5.53665868-0.50165471 i  &  5.55454509-0.50003931 i \\
10  & 10.51183964-0.50083853 i  & 10.51373778-0.50074811 i & 10.51943015-0.50047714 i & 10.52891067-0.50002650 i  \\
15  & 15.50804389-0.50038767 i & 15.50933215-0.50034610 i & 15.51319628-0.50022147 i & 15.51963438-0.50001396 i  \\
  \hline
\hline
$\ell$ &  $m b_0=0.8$ & $m b_0=1.0$ & $m b_0=1.2$ & $m b_0 = 1.4$  \\
%\colrule
\hline
0  & 0.95710482-0.43978143 i  & 1.10771470-0.37998677 i & 1.27720491-0.32956100 i & 1.45807345-0.28868020 i \\
1  & 1.74918582-0.47625873 i  & 1.84312741-0.45198450 i & 1.95313933-0.42652616 i & 2.07694738-0.40110069 i  \\
2   &  2.66507432-0.48988928 i  & 2.72972309-0.47828709 i & 2.80695095-0.46512795 i  & 2.89584111-0.45085047 i  \\
3  & 3.62200113-0.49464557 i  & 3.67048645-0.48811154 i & 3.72894874-0.48045895 i  & 3.79695454-0.47185363 i  \\
5  & 5.57949282-0.49780347 i & 5.61141062-0.49497195 i & 5.65018391-0.49157531 i & 5.69567673-0.48764897 i   \\
10 & 10.54216920-0.49939763 i & 10.55919163-0.49859255 i &  10.57995997-0.49761382 i & 10.60445239-0.49646451 i  \\
15  & 15.52864326-0.49972388 i & 15.54021845-0.49935166 i  & 15.55435426-0.49889785 i &   15.57104374-0.49836312 i \\
\hline
\end{tabular}
\end{table}

\begin{table}[h]
\centering
\caption{Fundamental $n=0$ QNFs $\omega b_0$ for scalar perturbations for the wormhole spacetime described by Eq. (\ref{wormhole}) with $M/b_0 =0.1$, for several values of the angular momentum and the mass of the scalar field using the CFM.}
\label{NVBE1}
\begin{tabular}{|c|c|c|c|c|}
\hline
$\ell$ &  $m b_0=0$ & $m b_0=0.2$ & $m b_0=0.4$ & $m b_0 = 0.6$ \\
%\colrule
\hline
0  & 0.64217472 - 0.52979187 i & 0.65556756 - 0.51733481 i  & 0.69692010 - 0.48206571 i  & 0.76830083 - 0.43049763 i  \\
1  & 1.42443104 - 0.44987061 i  & 1.43440781 - 0.44629804 i  & 1.46408863 - 0.43594442 i &  1.51272479 - 0.41981413 i \\
2   & 2.28995771 - 0.43248900 i & 2.29664353 - 0.43105279 i  & 2.31660278 - 0.42681157 i & 2.34954643 - 0.41995908 i  \\
3  & 3.17040105 - 0.42628276 i   & 3.17533511 - 0.42552688 i & 3.19009512 - 0.42327874 i & 3.21455621 - 0.41959546 i  \\
5  & 4.94550683 - 0.42182228 i & 4.94871268 - 0.42151025 i & 4.95831818 - 0.42057757 i  &  4.97428747 - 0.41903441 i \\
10  & 9.40544221 - 0.41934024 i  & 9.40713911 - 0.41925382 i & 9.41222797 - 0.41899482 i & 9.42070336 - 0.41856403 i  \\
15  & 13.87311667 - 0.41879955 i & 13.87426865 - 0.41875982 i & 13.87772405 - 0.41864067 i & 13.88348113 - 0.41844227 i  \\
  \hline
\hline
$\ell$ &  $m b_0=0.8$ & $m b_0=1.0$ & $m b_0=1.2$ & $m b_0 = 1.4$  \\
%\colrule
\hline
0  & 0.86912831 - 0.37246044 i  & 0.99422781 - 0.31703053 i & 1.13627138 - 0.26904009 i & 1.28910794 - 0.22942901 i \\
1  & 1.57908796 - 0.39932068 i  & 1.66154662 - 0.37599623 i & 1.75820800 - 0.35123892 i & 1.86709221 - 0.32616103 i  \\
2   &  2.39501097 - 0.41079193 i  & 2.45238427 - 0.39967609 i & 2.52093849 - 0.38701017 i  & 2.59986671 - 0.37319156 i  \\
3  & 3.24851550 - 0.41456774 i  & 3.29169959 - 0.40831378 i & 3.34377421- 0.40097186 i  & 3.40435536 - 0.39269243 i  \\
5  & 4.99656144 - 0.41689732 i & 5.02505886 - 0.41418882 i & 5.05967769 - 0.41093663 i & 5.10029687 - 0.40717299 i   \\
10 & 9.43255622 - 0.41796278 i & 9.44777395 - 0.41719290 i &  9.46634048 - 0.41625671 i & 9.48823630 - 0.41515702 i  \\
15  & 13.89153706 - 0.41816492 i & 13.90188784 - 0.41780899 i  & 13.91452839 - 0.41737499 i &   13.92945250 - 0.41686353 i \\
\hline
\end{tabular}
\end{table}
\begin{table}[h]
\centering
\caption{Fundamental $n=0$ QNFs $\omega b_0$ for scalar perturbations for the wormhole spacetime described by Eq. (\ref{wormhole}) with $M/b_0 =0.25$, for several values of the angular momentum and the mass of the scalar field using the CFM.}
\label{NVMT0}
\begin{tabular}{|c|c|c|c|c|}
\hline
$\ell$ &  $m b_0=0$ & $m b_0=0.2$ & $m b_0=0.4$ & $m b_0 = 0.6$ \\
%\colrule
\hline
0  & 0.57954862 - 0.38454705 i & 0.58788305 - 0.37466006 i & 0.61338274 - 0.34648046 i & 0.65699718 - 0.30486477 i  \\
1  & 1.17377910 - 0.30720290 i & 1.18130424 - 0.30383997 i & 1.20370378 - 0.29401135 i  &  1.24044319 - 0.27843471 i \\
2   & 1.84793711 - 0.28302527 i &  1.85314787 -0.28162482 i & 1.86870696 - 0.27747445 i &  1.89439869 - 0.27072045 i  \\
3  & 2.53780145 - 0.27199629 i  & 2.54168130 - 0.27124742 i & 2.55328871 - 0.26901575 i  &  2.57252849 - 0.26534505 i\\
5  & 3.93343488 - 0.26196975 i & 3.93596757 - 0.26165516 i & 3.94355629 - 0.26071402 i &  3.95617323 - 0.25915412 i \\
10  & 7.45015539 - 0.25438498 i & 7.45149755 - 0.25429618 i  & 7.45552258 - 0.25402996 i &  7.46222623 - 0.25358694 i \\
15  & 10.97800980 - 0.25223172 i & 10.97892087 - 0.25219058 i & 10.98165362 - 0.25206720 i  &  10.98620671 - 0.25186172 i \\
  \hline
\hline
$\ell$ &  $m b_0=0.8$ & $m b_0=1.0$ & $m b_0=1.2$ & $m b_0 = 1.4$  \\
%\colrule
\hline
0  & 0.71935825 - 0.25784028 i & 0.80016054 - 0.21267941 i & 0.89645704 - 0.17253589 i & 1.00377455 - 0.13801622 i \\
1  & 1.29062276 - 0.25811736 i & 1.35299556 - 0.23413835 i & 1.42601082 - 0.20745244 i  &  1.50784347 - 0.17879248 i \\
2   & 1.92987630 - 0.26158567 i & 1.97468010 - 0.25034262 i & 2.02826070 - 0.23728299 i &   2.09000500 - 0.22268827 i\\
3  &  2.59924592 - 0.26030455 i  & 2.63323242 - 0.25398401 i   & 2.67423284 - 0.24648753 i & 2.72195391 - 0.23792684 i \\
5  & 3.97377260 - 0.25698816 i & 3.99629142 - 0.25423331 i & 4.02365065 - 0.25091069 i &  4.05575648 - 0.24704466 i  \\
10  & 7.47160139 - 0.25296810 i & 7.48363819 - 0.25217484 i & 7.49832401 - 0.25120891 i &  7.51564357 - 0.25007240 i \\
15  & 10.99257790 - 0.25157433 i & 11.00076404 - 0.25120533 i & 11.01076114 - 0.25075510 i &   11.02256429 - 0.25022409 i\\
\hline
\end{tabular}
\end{table}

\begin{table}[h]
\centering
\caption{First overtone $n=1$ QNFs $\omega b_0$ for scalar perturbations for the Bronnikov-Ellis wormhole for several values of the angular momentum and the mass of the scalar field using the CFM.}
\label{TableIV}
\begin{tabular}{|c|c|c|c|c|}
\hline
$\ell$ &  $m b_0=0$ & $m b_0=0.2$ & $m b_0=0.4$ & $m b_0 = 0.6$ \\
%\colrule
\hline
0  & 0.46716556-2.17646191 i & 0.46906191-2.16766276 i  & 0.47488384-2.14108791 i  & 0.48505482-2.09619201 i  \\
1  & 1.25582759-1.70245148 i  & 1.26145961-1.69485057 i  & 1.27859466-1.67213708 i &  1.30794582-1.63461322 i \\
2   & 2.34497715-1.57249831 i & 2.35086218-1.56856180 i  & 2.36853730-1.55685647 i & 2.39805487-1.53769317 i  \\
3  & 3.39009976-1.53721055 i & 3.39499204-1.53499538 i & 3.40965523-1.52839415 i & 3.43404739-1.51753791 i  \\
5  & 5.43090081-1.51524731 i & 5.43431671-1.51429486 i & 5.44455525-1.51144721 i  &  5.46158899-1.50673327 i \\
10  & 10.46413347-1.50422829 i  & 10.46600591-1.50395918 i & 10.47162139-1.50315267 i & 10.48097437-1.50181129 i  \\
15  & 15.47575755-1.50194589 i & 15.47703778-1.50182166 i & 15.48087787-1.50144912 i & 15.48727599-1.50082884 i  \\
  \hline
\hline
$\ell$ &  $m b_0=0.8$ & $m b_0=1.0$ & $m b_0=1.2$ & $m b_0 = 1.4$  \\
%\colrule
\hline
0  & 0.50037244-2.03202248 i  & 0.52218113-1.94715585 i & 0.55270365-1.83962607 i & 0.59567461-1.70691857 i \\
1  & 1.35067753-1.58289857 i  & 1.40832928-1.51810061 i & 1.48262974-1.44202256 i & 1.57514763-1.35732390 i  \\
2   &  2.43947742-1.51158301 i  & 2.49284354-1.47922345 i & 2.55812690-1.44147368 i  & 2.63519525-1.39931666 i  \\
3  & 3.46809539-1.50263950 i  & 3.51169079-1.48398519 i & 3.56468519-1.46192352 i  & 3.62688609-1.43685161 i  \\
5  & 5.48537240-1.50020039 i & 5.51584212-1.49191323 i & 5.55291737-1.48195215 i & 5.59650057-1.47041133 i   \\
10 & 10.49405566-1.49993922 i & 10.51085246-1.49754225 i &  10.53134838-1.49462776 i & 10.55552357-1.49120463 i  \\
15  & 15.49622907-1.49996173 i & 15.50773287-1.49884904 i  & 15.52178191-1.49749240 i &   15.53836956-1.49589379 i \\
\hline
\end{tabular}
\end{table}

\begin{table}[h]
\centering
\caption{First overtone $n=1$ QNFs $\omega b_0$ for scalar perturbations for the wormhole spacetime described by Eq. (\ref{wormhole}) with $M/b_0 =0.1$, for several values of the angular momentum and the mass of the scalar field using the CFM.}
\label{NVBE12}
\begin{tabular}{|c|c|c|c|c|}
\hline
$\ell$ &  $m b_0=0$ & $m b_0=0.2$ & $m b_0=0.4$ & $m b_0 = 0.6$ \\
%\colrule
\hline
0  & 0.48215901 - 1.83934960 i  & 0.48355134 - 1.83211141 i & 0.48784214 - 1.81024999 i  & 0.49539239 - 1.77331563 i    \\
1  & 1.18531283 - 1.43535637 i & 1.18973778 - 1.42849651 i & 1.20321550 - 1.40798383 i  & 1.22635308 - 1.37404779 i  \\
2   & 2.13978782 - 1.32436578 i  & 2.14484682 - 1.32066795 i & 2.16004435 - 1.30966487 i &  2.18543533 - 1.29162649 i \\
3  &  3.06406448 - 1.29198825 i & 3.06835811 - 1.28988826 i  & 3.08122789 - 1.28362767 i & 3.10263958 - 1.27332322 i  \\
5  & 4.87885201 - 1.27072447 i & 4.88188492 - 1.26981683 i & 4.89097566 - 1.26710268 i &  4.90610018 - 1.26260811 i \\
10  & 9.37098143 - 1.25947857 i & 9.37265290 - 1.25922149 i  & 9.37766566 - 1.25845099 i  & 9.38601480 - 1.25716938 i  \\
15  & 13.84985433 - 1.25707137 i  & 13.85099838 - 1.25695263 i & 13.85442998 - 1.25659656 i &  13.86014749 - 1.25600367 i \\
  \hline
\hline
$\ell$ &  $m b_0=0.8$ & $m b_0=1.0$ & $m b_0=1.2$ & $m b_0 = 1.4$  \\
%\colrule
\hline
0  & 0.50687537 - 1.72052831 i & 0.52341091 - 1.65073531 i & 0.54681941 - 1.56236912 i & 0.58009574 - 1.45346680 i  \\
1  & 1.26014835 - 1.32716772 i & 1.30593642 - 1.26821760 i & 1.36524947 - 1.19864836 i & 1.43954101 - 1.12064318 i  \\
2   & 2.22108946 - 1.26699755 i  & 2.26706119 - 1.23638676 i & 2.32335302 - 1.20054689 i &  2.38987830 - 1.16034332 i \\
3  & 3.13253313 - 1.25916454 i  & 3.17081881 - 1.24140704 i & 3.21737287 - 1.22036208 i & 3.27203347 - 1.19638504 i \\
5  & 4.92721862 - 1.25637588 i  & 4.60133237 - 2.98594631 i & 4.62635318 - 2.96656927 i &  5.02590622 - 1.22790668 i \\
10  & 9.39769212 - 1.25538044 i & 9.41268624 - 1.25308945 i & 9.43098258 - 1.25030310 i & 9.45256345 - 1.24702947 i  \\
15  & 13.86814819 - 1.25517478 i & 13.87842828 - 1.25411105 i & 13.89098288 - 1.25281395 i & 13.90580606 - 1.25128527 i  \\
\hline
\end{tabular}
\end{table}
\begin{table}[h]
\centering
\caption{First overtone $n=1$ QNFs $\omega b_0$ for scalar perturbations for the wormhole spacetime described by Eq. (\ref{wormhole}) with $M/b_0 =0.25$, for several values of the angular momentum and the mass of the scalar field using the CFM.}
\label{NVMT02}
\begin{tabular}{|c|c|c|c|c|}
\hline
$\ell$ &  $m b_0=0$ & $m b_0=0.2$ & $m b_0=0.4$ & $m b_0 = 0.6$ \\
%\colrule
\hline
0  & 0.58217648 - 1.31200598 i & 0.58178419 - 1.30705341 i & 0.58074738 - 1.29211492 i & 0.57947867 - 1.26692340 i \\
1  & 1.10576623 - 0.99897774 i  & 1.10818210 - 0.99270181 i & 1.11554218 - 0.97388846 i & 1.12818333 - 0.94260457 i  \\
2   & 1.82380452 - 0.88934468 i & 1.82753835 - 0.88579921 i  & 1.83876095 - 0.87522321 i & 1.85753019 - 0.85779732 i  \\
3  &  2.53427660 - 0.84472565 i  & 2.53759327 - 0.84267946 i & 2.54753715 - 0.83656988 i &  2.56408904 - 0.82648272 i \\
5  & 3.94453789 - 0.80453084 i & 3.94692945 - 0.80363037 i & 3.95409834 - 0.80093568 i & 3.96602715 - 0.79646673 i  \\
10  & 7.46522687 - 0.77201761 i & 7.46655166 - 0.77175658 i & 7.47052476 - 0.77097406 i & 7.47714240 - 0.76967182 i  \\
15  & 10.99099654 - 0.76178973 i  & 10.99190286 - 0.76166781 i & 10.99462141 - 0.76130218 i &  10.99915090 - 0.76069321 i \\
  \hline
\hline
$\ell$ &  $m b_0=0.8$ & $m b_0=1.0$ & $m b_0=1.2$ & $m b_0 = 1.4$  \\
%\colrule
\hline
0  & 0.57864581 - 1.23096746 i & 0.57916915 - 1.18343640 i & 0.58228934 - 1.12319399 i & 0.58978003 - 1.04879086 i   \\
1  & 1.14666028 - 0.89904448 i  & 1.17171924 - 0.84366508 i & 1.20422868 - 0.77738301 i & 1.24504510 - 0.70180079 i  \\
2   & 1.88392504 - 0.83381987 i & 1.91802066 - 0.80370035 i & 1.95985689 - 0.76794601 i & 2.00940382 - 0.72713929 i  \\
3  & 2.58721432- 0.81255726 i  & 2.61685953 - 0.79498064 i & 2.65294831 - 0.77398025 i & 2.69537748 - 0.74981414 i \\
5  & 3.98268698 - 0.79025619 i  & 4.00403759 - 0.78234861 i & 4.03002761 - 0.77279927 i & 4.06059492 - 0.76167277 i  \\
10  & 7.48639830 - 0.76785275 i & 7.49828374 - 0.76552089 i & 7.51278752 - 0.76268135 i &  7.52989613 - 0.75934031 i \\
15  & 11.00548923 - 0.75984153 i & 11.01363344 - 0.75874802 & 11.02357975 - 0.75741379 i  &  11.03532354 - 0.75584020 i \\
\hline
\end{tabular}
\end{table}

%\clearpage

\section{Comparison of CFM and WKB method}
\label{AFC}

In Table \ref{TableCom} we show the fundamental QNFs and the first overtone calculated using the CFM and the WKB approximation, and we see a good agreement for high $\ell$. We show the relative error of the real and imaginary parts of the values obtained with the WKB method with respect to the values obtained with the CFM, which is defined by
\begin{eqnarray}
\label{E}
\epsilon_{Re(\omega)} &=& \frac{\mid Re(\omega_{WKB})- Re(\omega_{CFM})\mid}{\mid Re(\omega_{CFM}) \mid}\cdot 100\%\,, \\
\epsilon_{Im(\omega)} &=&\frac{\mid Im(\omega_{WKB}) - Im(\omega_{CFM})\mid}{\mid Im(\omega_{CFM})\mid } \cdot 100\% \,,
\end{eqnarray}
where $\omega_{WKB}$ corresponds to the result obtained with the six-order WKB method, and $\omega_{CFM}$ denotes the result with the CFM.  We can observe that the error does not exceed  $\approx$ 8.819 $\%$ in the imaginary part, and   $\approx$ 2.309 $\%$ in the real part, for low values of $\ell$. However, for high values of $\ell$ ($\ell = 10, 15$), the error does not exceed 0.0007 $\%$ in the imaginary part, and  3.60 $\cdot 10^{-5}\, \%$ in the real part. Also, as it was observed, the frequencies all have a negative imaginary part, which means that the propagation of massive scalar fields is stable in this background.

%We observe that when $M$ increases the WKB method is less accurate for low values of $\ell$.

%\clearpage

\begin{table}[H]
\centering
\caption{QNFs $\omega b_0$ for massive scalar field perturbations $m=0.5$ for the wormhole spacetime described by Eq. (\ref{wormhole}) with $M/b_0=0.1$, $n=0,1$, for several values of the angular momentum $\ell$ of the scalar field using the CFM and the six-order WKB approximation.}
\begin{tabular}{|c|c|c|c|c|}  \hline
\multicolumn{5}{|c|}{$n=0$}  \\  \hline
$\ell$ &  CFM & WKB & $\epsilon_{Re(\tilde{\omega})}(\%)$ & $\epsilon_{Im(\tilde{\omega})}(\%)$ \\
%\colrule
\hline
 % 0  &  0.72878751 - 0.45774669 i & 0.76209632 - 0.35470112 i & 4.5704 & 22.5115 \\
  1  &  1.48610337 - 0.42851924 i  & 1.50406001 - 0.41070253 i & 1.2083  & 4.1577 \\
  2  & 2.33147670 - 0.42369524 i & 2.33358638 - 0.42204168 i & 0.0905 & 0.3903 \\
  3   & 3.20112396 - 0.42161155 i &  3.20146777 - 0.42132985 i   & 0.0107 & 0.0668  \\
    5  & 4.96551051 - 0.41988141 i & 4.96553585 - 0.41985314 i & 0.0005 &  0.0067 \\

   10  & 9.41604283 - 0.41880083 i  & 9.41604328 - 0.41879991 i & $4.78 \cdot 10^{-6}$ & 0.0002 \\
  15  & 13.88031503 - 0.41855136 i  & 13.88031506 - 0.41855126 i  &  $2.16 \cdot 10^{-7}$ & $2.39 \cdot 10^{-5}$ \\
\hline
\multicolumn{5}{|c|}{$n=1$} \\ \hline
$\ell$ &  CFM & WKB & $\epsilon_{Re(\tilde{\omega})}(\%)$ & $\epsilon_{Im(\tilde{\omega})}(\%)$ \\
%\colrule
\hline
 % 0  &  0.49117622 - 1.79370750  i & 0.41768809 - 1.55508806 i & 14.9617 & 13.3031 \\
  1  &   1.21352327 - 1.39267106 i  &  1.24154163 - 1.26985185 i  & 2.3088 & 8.8190 \\
  2  & 2.17146134 - 1.30150131  i & 2.18112283 - 1.28436587 i & 0.4449 & 1.3166  \\
  3   &  3.09086911 - 1.27897071 i &  3.09292339 - 1.27587774 i &  0.0665 & 0.2418 \\
    5  & 4.89778582 - 1.26507565 i  & 4.89796016 - 1.26478685 i & 0.0036 & 0.0228  \\
   10  & 9.38142361 - 1.25787387 i  & 9.38142698 - 1.25786520 i & $3.60 \cdot 10^{-5}$ & 0.0007 \\
  15  & 13.85700314 - 1.25632967 i & 13.85700340 - 1.25632873 i & $1.88 \cdot 10^{-6}$ & $7.48 \cdot 10^{-5}$ \\
\hline

\end{tabular}
\label{TableCom}
\end{table}

%\section{WKB method}
%\label{WKB}

%In this appendix we show

\end{document}